SIMPLE ALTERNATIVES TO THE COMMON CORRELATED EFFECTS MODEL

Nicholas L. Brown
Michigan State University

Peter Schmidt
Michigan State University

Jeffrey M. Wooldridge
Michigan State University

**Abstract**: We study estimation of factor models in a fixed-$T$ panel data setting and significantly relax the common correlated effects (CCE) assumptions pioneered by Pesaran (2006) and used in dozens of papers since. In the simplest case, we model the unobserved factors as functions of the cross-sectional averages of the explanatory variables and show that this is implied by Pesaran's assumptions when the number of factors does not exceed the number of explanatory variables. Our approach allows discrete explanatory variables and flexible functional forms in the covariates. Plus, it extends to a framework that easily incorporates general functions of cross-sectional moments, in addition to heterogeneous intercepts and time trends. Our proposed estimators include Pesaran's pooled correlated common effects (CCEP) estimator as a special case. We also show that in the presence of heterogeneous slopes our estimator is consistent under assumptions much weaker than those previously used. We derive the fixed-$T$ asymptotic normality of a general estimator and show how to adjust for estimation of the population moments in the factor loading equation.

# 1. Introduction

The common correlated effects (CCE) approach to linear panel data models with common factors has become very influential both in the theoretical panel data literature and in empirical applications. Introduced by Pesaran (2006), the CCE model includes an equation of interest, in which the error has a factor structure, and a reduced form equation in which the explanatory variables are linear functions of the same factors appearing in the main equation. CCE then treats the cross-sectional averages of the response and explanatory variables as fixed effects, eliminating (asymptotically) the unobserved heterogeneity. Consistency and asymptotic normality of the pooled CCE (CCEP) estimator was originally proved for the case of homogeneous slopes when both the number of units, $N$, and the number of time periods, $T$, tend to infinity such that $T/N \to 0$.

Since Pesaran's pioneering work, pooled CCE estimation and related estimators have been has been extensively studied in the context of large-$T$ panels. Theoretical papers that examine both robustness properties and extensions of CCE estimation, with $T \to \infty$, include Chudik, Pesaran, and Tosetti (2011), Westerlund and Urbain (2013, 2015), Chudik and Pesaran (2015), Neal (2015), Karabiyik, Reese, and Westerlund (2017) [hereafter, KRW (2017)], Westerlund (2018), Chen and Yan (2019), De Vos and Westerlund (2019), Karabiyik, Urbain, and Westerlund (2019), De Vos and Everaert (2021), Norkute, Sarafidis, Yamagata, and Cui (2021), and Kapetanios, Serlenga, and Shin (2021).

Despite positive results in simulation studies from Chudik, Pesaran, and Tosetti (2011), Chudik and Pesaran (2015), Westerlund and Urbain (2015), and Breitung and Hansen (2020), comparatively less is known about the fixed-$T$ theoretical properties of CCE. Pesaran (2006) showed that the mean group CCE estimator is consistent for fixed $T$ but made strong rank and independence assumptions. Pesaran's (2006) asymptotic normality result requires $T \to \infty$ along with $N$. Westerlund, Petrova, and Norkute (2019) [hereafter, WPN (2019)] proved fixed-$T$ asymptotic normality of the CCEP. De Vos and Everaert (2021) provide an analytic inconsistency correction for CCEP in dynamic models when $T$ is fixed, but asymptotic normality requires $T$ to grow to infinity. Collectively, these papers open the door for microeconometric applications of CCEP but they still make stringent restrictions on the stochastic components of the model. First, they assume a factor structure for the explanatory variables, ruling out many staples of microeconometric applications such as dummy explanatory variables and interaction effects. Second, the factor structure is assumed independent of all idiosyncratic errors, as well as the loadings associated with the factor structure of the covariates. Third, the assumptions are very restrictive concerning the time dependence and stationarity of the idiosyncratic errors over time. Fourth, the current fixed-$T$ literature assumes homogeneity of the slope coefficients – something not assumed in Pesaran (2006). In this paper we substantially relax all of these assumptions in a constant coefficients model and also study CCEP and extensions in a model with heterogeneous slopes.

For the purposes of this paper, there is another important feature of current approaches to common factor models. Namely, if the covariates satisfy a pure factor structure as in Pesaran (2006) and WPN (2019), then the $T \times k$ matrix of cross-sectional expectations, $\boldsymbol{\mu}_X$, is generally rank-deficient: $\text{rank}(\boldsymbol{\mu}_X) < k$. Due to this degeneracy, the asymptotic analysis of CCE estimators is complicated and cumbersome; see KRW (2017). Aside from the difficulties it implies for asymptotic analysis, we find the degeneracy unappealing. Conceptually, even with somewhat small $T$, there is no reason to think the cross sectional averages of the covariates would be perfectly linearly related. Moreover, deficient rank is not seen in microeconometric

applications – something we explore further in Section 2. Instead, we propose an alternative assumption in the context of a linear factor model, which is that the factors are linear functions of $\boldsymbol{\mu}_X$ and possibly the $T \times 1$ vector of cross-sectional means of the dependent variable, $\boldsymbol{\mu}_y$. We show that Pesaran's CCE setup implies our preferred assumption. Unlike Pesaran (2006) and WPN (2019), $\boldsymbol{\mu}_X$ always has full rank. Our representation is much easier to work with and more realistic. We can easily handle situations where nonlinear functions of explanatory variables appear, or discrete explanatory variables, without change. We can incorporate other functions of cross-sectional means and even other cross-sectional moments. Because we need not worry about deficient rank of $\boldsymbol{\mu}_X$ we are only limited by how large $T$ is relative to the number of moments we include to proxy for the unobserved factors. As in Pesaran (2006), we can also include standard forms of heterogeneity, such as unit-specific intercepts and time trends.

To summarize, in this paper we provide a framework for applying the CCEP estimator, and extensions, to panels where the number of time periods is small (fixed in the asymptotic analysis) compared to the number of cross-sectional units. We significantly relax the restrictions on the stochastic elements of the model. Our factor structure allows linear and nonlinear functions of cross-sectional means and other moments as well as unobserved effects or unit-specific trends in addition to cross-sectional averages. Consequently, the framework provides a unification and extension of Wooldridge (2005) and Pesaran (2006) in the fixed-$T$ setting. We also derive conditions under which the mean value of heterogeneous slopes can be consistently estimated. Our asymptotic normality result and asymptotic variance matrix estimator are straightforward and apply to all cases without modification. A key contribution is that we explicitly address the sampling variation that arises from estimating the cross-sectional moments. We also clarify the role of observed, deterministic variables in the equation of interest and show that any variables that change only across time can be ignored provided the extended CCEP estimator includes the cross-sectional averages of the explanatory variables.

The remainder of the paper is organized as follows. In Section 2 we introduce the basic model proposed by Pesaran (2006), discuss the CCE structure, and discuss alternative identification assumptions. We show that, under the fixed-$T$ assumptions used in WPN (2019), the unobserved factors can be written as linear functions of the cross-sectional means of the explanatory variables or as linear functions of the cross-sectional means of the explanatory variables and the response variable. In Section 3 we impose a full rank condition on the matrix of cross-sectional means and sketch consistency of the CCEP estimator. We study the consequences of adding known deterministic factors in Section 4. Interestingly, whenever the cross-sectional averages of the explanatory variables are included in any CCEP-type estimator the estimated coefficients on the explanatory variables do not change. This appears to be a novel result. In Section 5 we provide a more general framework where the factors can be related to estimable functions of the moments beyond linear functions of the cross sectional means. We also allow deterministic factors. In this general framework we formally prove consistency of an extended CCEP (ECCEP) estimator. Section 6 extends the consistency of the ECCEP estimator when the slopes are heterogeneous, using a generalization of an assumption in Wooldridge (2005). In Section 7 we derive a fixed-$T$ asymptotic normality result for the ECCEP estimator that applies to the basic model and the model with heterogeneous slopes. Section 8 provides a simple, consistent estimator of the asymptotic variance. For our asymptotic analysis we assume random sampling in the cross section with fixed $T$ and $N \to \infty$. Therefore, the asymptotic theory is standard, and so we do not state moment conditions underlying the law of large numbers and central limit theorem.

In Section 9 we discuss some practical aspects of implementing CCEP, including the decision to include only the cross-sectional averages of the explanatory variables or adding the averages of the outcome variable. Section 10 contains concluding remarks.

## 2. **The Basic Model with Constant Slopes**

We begin with a model having homogeneous slope coefficients, as in WPN (2019). For a random draw $i$ from the cross section,

$$y_{it} = \mathbf{x}_{it}\boldsymbol{\beta} + \mathbf{f}_t\boldsymbol{\gamma}_i + e_{it}, \, t = 1,\ldots,T \tag{2.1}$$

The fixed $k \times 1$ vector of parameters, $\boldsymbol{\beta}$, is of interest. The factors in the $1 \times p$ row vector $\mathbf{f}_t$ are not random but they are unobserved; therefore, $p$ is not known, either. Robertson and Sarafidis (2015) also take the $\mathbf{f}_t$ to be nonrandom. Other fixed-$T$ approaches, particularly WPN (2019), view the $\mathbf{f}_t$ as random variables. In the fixed-$T$ setting the results are typically obtained by first conditioning on $\{\mathbf{f}_t : t = 1,\ldots,T\}$, resulting in the same inference as treating the $\mathbf{f}_t$ as nonrandom. Our view is that the $\mathbf{f}_t$ play the same role as an intercept and linear trend in a heterogenous trend model; the difference is that we do not directly observe $\mathbf{f}_t$.

All other variables in equation (2.1), including the factor loadings, $\boldsymbol{\gamma}_i$, are random. We observe only $y_{it}$ and $\mathbf{x}_{it}$. The vector $\mathbf{x}_{it}$ does not include known, deterministic factors such as an overall intercept or time period dummies. In Section 4 we discuss including such variables.

The $\{e_{it} : t = 1,\ldots,T\}$ are the idiosyncratic errors. Many treatments of the unobserved effects model in (2.1) impose strong restrictions on the stationarity and dependence properties of $\{e_{it} : t = 1,\ldots,T\}$. Other than finite moment conditions, we will impose no substantive assumptions in this paper.

Stacking across $t$ for a cross-sectional unit $i$ we can write

$$\mathbf{y}_i = \mathbf{X}_i \boldsymbol{\beta} + \mathbf{F}\boldsymbol{\gamma}_i + \mathbf{e}_i, \qquad (2.2)$$

where $\mathbf{y}_i$ is $T \times 1$, $\mathbf{X}_i$ is $T \times k$, $\mathbf{F}$ is $T \times p$, and $\mathbf{e}_i$ is $T \times 1$. The factor literature assumes that rank($\mathbf{F}$) = $p$, which simply means there are no redundant factors. Below we propose a setting where the value of $p$ and the rank of $\mathbf{F}$ are essentially irrelevant.

In the large-$T$ case, Pesaran (2006) added a factor structure to the matrix of regressors:

$$\mathbf{X}_i = \mathbf{F}\boldsymbol{\Gamma}_i + \mathbf{V}_i, \qquad (2.3)$$

so that $\mathbf{X}_i$ depends linearly on the same factors that appear in equation (2.2). Taken together, (2.2) and (2.3) constitute a "common correlated effects" (CCE) model in the terminology of Pesaran (2006). The equations in (2.3) are not of particular interest – they act as a kind of reduced form – but Pesaran added them in the large-$T$ case in order to estimate $\boldsymbol{\beta}$. The same structure has been used by WPN (2019) in the fixed-$T$ case. We call (2.3) the "strong common factor" (SCF) assumption.

Combined with assumptions about the elements of $\boldsymbol{\Gamma}_i$ and $\mathbf{V}_i$ that are usually imposed, the SCF assumption is very restrictive. For example, Pesaran (2006) assumes $\boldsymbol{\Gamma}_i$ and $\mathbf{V}_i$ are independent, which essentially rules out discreteness in the elements of $\mathbf{X}_i$. This precludes any policy analysis defined by discrete interventions. Moreover, as discussed by De Vos and Westerlund (2019), (2.3) cannot be justified when the covariates include functional forms – such as squares and interactions of underlying explanatory variables – that are staples of applied econometrics. Independence of the $\mathbf{v}_{it}$ across $t$ is also commonly assumed, implying that all serial dependence in $\mathbf{x}_{it}$ is due to the heterogeneity, $\boldsymbol{\Gamma}_i$. Again, this is a strong assumption that seems unlikely to be true. For example, if the $\mathbf{x}_{it}$ are inputs in a production function it seems unrealistic that inputs for a given firm would be independent over time. Nevertheless, SCF has been used even in the recent small-$T$ factor model literature; see, for example, WPN (2019).

Pesaran (2006) ($T \to \infty$) and WPN (2019) (fixed $T$) assume $p \leq k+1$ in (2.2) and (2.3). In this section we assume $p \leq k$, although we relax this later. Most microeconometric studies include several covariates; consequently, this restriction on the number of factors still allows for substantial heterogeneity. The standard model with a single additive heterogeneity has $\mathbf{f}_t = 1$ and so $p = 1$.

If we make the traditional, unrestrictive assumption $E(\mathbf{V}_i) = \mathbf{0}$, SCF implies

$$E(\mathbf{X}_i) = \mathbf{F} E(\mathbf{\Gamma}_i) \equiv \mathbf{F}\mathbf{\Gamma}, \tag{2.4}$$

where $\mathbf{\Gamma}$ is a $p \times k$ nonrandom matrix with rank$(\mathbf{\Gamma}) = p \leq k$. Note that the zero mean assumption on $\mathbf{V}_i$ imposes no assumptions on the relationship between $\mathbf{V}_i$ and $\mathbf{\Gamma}_i$ or among any of the unobservables in (2.2) and (2.3). Let $\boldsymbol{\mu}_\mathbf{X}$ be the $T \times k$ matrix of means,

$$\boldsymbol{\mu}_\mathbf{X} \equiv E(\mathbf{X}_i), \tag{2.5}$$

so that the $t^{th}$ row of $\boldsymbol{\mu}_\mathbf{X}$ is $\boldsymbol{\mu}_t^\mathbf{x} \equiv E(\mathbf{x}_{it})$. Then we can write (2.5) as

$$\boldsymbol{\mu}_\mathbf{X} = \mathbf{F}\mathbf{\Gamma} \tag{2.6}$$

We call (2.6) the "weak common factor" (WCF) assumption because it imposes none of the strong restrictions typically imposed on (2.3).

Even the WCF assumption is unappealing. For one, it implies that if $p < k$ then $\boldsymbol{\mu}_\mathbf{X}$ has rank less than $k$, a peculiar restriction that fails in basic models and has little, if any, empirical support. As a simple example, take $T = 3$, $\mathbf{f}_t = 1$ (so $p = 1$), and assume we have a staggered policy intervention, with $t = 1$ denoting a control period where no units are subjected to the intervention (or "treatment"). Let $x_{it1}$ and $x_{it2}$ denote the binary intervention indicators, where some units are treated in periods two and three (indicated by $x_{it1} = 1, t = 2, 3$) and others only in period three ($x_{it2} = 1, t = 3$). Then

$$\boldsymbol{\mu}_{\mathbf{X}} = \begin{pmatrix} 0 & 0 \\ \rho_1 & 0 \\ \rho_1 & \rho_2 \end{pmatrix}$$

where $\rho_1$ is the fraction of population units subjected to the intervention in the early period and $\rho_2$ is the fraction of treated units in the last period. In no reasonable scenario does $\boldsymbol{\mu}_{\mathbf{X}}$ have rank less than two, and yet this is implied by (2.6) if there is only a single factor.

If $p = k$ in (2.6) and we make the standard assumption rank($\mathbf{F}$) = $p$ then $\boldsymbol{\mu}_{\mathbf{X}}$ has rank $k$ and the deficient rank implication of WCF (and SCF) disappears. Moreover, we can write $\mathbf{F} = \boldsymbol{\mu}_{\mathbf{X}} \boldsymbol{\Gamma}^{-1}$, so the factors are a nonsingular linear combinations of the means. Our proposal is to replace WCF with an assumption that models the factors as linear functions of $E(\mathbf{X}_i)$ even when $p < k$:

$$\mathbf{F} = \boldsymbol{\mu}_{\mathbf{X}} \boldsymbol{\Lambda}_{\mathbf{X}}, \tag{2.7}$$

where $\boldsymbol{\Lambda}_{\mathbf{X}}$ is a $k \times p$ matrix of unknown parameters. We (somewhat immodestly) refer to this as assumption BSW. Importantly, if we start with (2.7) then we do not need to know or estimate the number of factors, $p$, and we need not restrict rank($\mathbf{F}$).

When $p = k$, WCF [with rank($\boldsymbol{\Gamma}$) = $p$] and BSW are clearly equivalent. When $p < k$, BSW is strictly weaker than WCF. To see this, assume (2.6). Because $\boldsymbol{\Gamma}$ has rank $p$, we can partition $\boldsymbol{\Gamma}$ as $(\boldsymbol{\Gamma}_1 | \boldsymbol{\Gamma}_2)$ where $\boldsymbol{\Gamma}_1$ is $p \times p$ and nonsingular. Therefore, $\boldsymbol{\mu}_{\mathbf{X}} = (\mathbf{F}\boldsymbol{\Gamma}_1 | \mathbf{F}\boldsymbol{\Gamma}_2)$ and we can write

$$\mathbf{F} = \boldsymbol{\mu}_{\mathbf{X}}\boldsymbol{\Lambda}_{\mathbf{X}}, \ \boldsymbol{\Lambda}_{\mathbf{X}} \equiv \begin{pmatrix} \boldsymbol{\Gamma}_1^{-1} \\ \mathbf{0} \end{pmatrix}.$$

[Clearly rank($\boldsymbol{\mu}_{\mathbf{X}}$) < $k$ when $p < k$.] To see that BSW does not imply WCF, note that WCF requires $\boldsymbol{\mu}_{\mathbf{X}}$ to have rank less than $k$ when $p < k$ and BSW has no such implication. (BSW does not even require $\boldsymbol{\Lambda}_{\mathbf{X}}$ to have full rank.) If we start with (2.7) and $p < k$, $\boldsymbol{\mu}_{\mathbf{X}}$ may or may not have full column rank and $\mathbf{F}$ may or may not have full column rank. Our view is that assuming rank($\boldsymbol{\mu}_{\mathbf{X}}$) = $k$ is more realistic than the WCF implication that rank($\boldsymbol{\mu}_{\mathbf{X}}$) < $k$.

If we start with (2.7) and add the empirically relevant assumption rank($\boldsymbol{\mu}_{\mathbf{X}}$) = $k$ then there is no reason to consider the possibility that $p > k$. If (2.7) holds then rank($\mathbf{F}$) $\leq k$. Suppose that $\mathbf{F}$ has the largest possible rank, rank($\mathbf{F}$) = $k$. Then we can write write $\mathbf{F} = (\mathbf{F}_1, \mathbf{F}_2)$ where $\mathbf{F}_1$ is $T \times k$ with rank($\mathbf{F}_1$) = $k$ and $\mathbf{F}_2 = \mathbf{F}_1 \mathbf{A}$ for some $\mathbf{A}$ of dimension $T \times (p - k)$. It follows that

$$\mathbf{F}\boldsymbol{\gamma}_i = \mathbf{F}_1\boldsymbol{\gamma}_{i1} + \mathbf{F}_2\boldsymbol{\gamma}_{i2} = \mathbf{F}_1\boldsymbol{\gamma}_{i1} + \mathbf{F}_1\mathbf{A}\boldsymbol{\gamma}_{i2} = \mathbf{F}_1(\boldsymbol{\gamma}_{i1} + \mathbf{A}\boldsymbol{\gamma}_{i2}), \tag{2.8}$$

which is a factor structure with $p = k$.

To summarize, we have shown that (2.6) and (2.7) are equivalent when $p = k$ and the BSW assumption (2.7) is strictly weaker when $p < k$. In addition to being weaker than WCF assumptions – and much weaker than implementations of SCF – there are other substantial benefits to starting with the representation in (2.7). For one, it allows augmenting the means $\boldsymbol{\mu}_t^x$ with observed factors that are staples of the microeconometric panel data literature. One simple, attractive extension is to define a row vector as

$$\boldsymbol{\psi}_t = (1, \boldsymbol{\mu}_t^x) \tag{2.9}$$

so that a standard additive effect, $\gamma_{i1}$, appears in (2.1). Provided $T$ is sufficiently large, we can add a linear trend to the factors as $\boldsymbol{\psi}_t = (1, t, \boldsymbol{\mu}_t^x)$, which leads to a heterogeneous (or "random") trend model. These possibilities are allowed starting with Pesaran (2006) but under SCF or similarly strong assumptions. As mentioned earlier, such assumptions make the asymptotic analysis of CCEP estimators, with or without observed factors, very difficult; in our view, unnecessarily so.

In Section 5 we study a general extension that includes (2.7) by assuming

$$\mathbf{F} = \mathbf{\Psi}\mathbf{\Lambda}, \tag{2.10}$$

where $\mathbf{\Psi}$ is a $T \times m$ matrix of rank $m$ that includes $\boldsymbol{\mu}_t^x$, possibly $\boldsymbol{\mu}_t^y$, known factors, and possible other unknown moments.

## 3. CCEP Estimation in the Model with Constant Slopes

We now turn to estimation of $\boldsymbol{\beta}$ in the context of equation (2.2). We do not include observed deterministic variables – such as an overall intercept, time trends, or time dummies – in the estimation. In Section 4 we show that including such variables does not affect estimation of $\boldsymbol{\beta}$. For the asymptotic derivations in this section, which are purposefully informal, we assume random sampling in the cross-sectional dimension and assume that all necessary moments are finite. A formal, more general, result is given in Section 5.

### 3.1. CCEP using Covariate Averages

Given the discussion in Section 2, it is natural to consider estimation of $\boldsymbol{\beta}$ in (2.2) under (2.7) when $E(\mathbf{X}_i)$ has full column rank. Plugging (2.7) into (2.2) gives

$$\mathbf{y}_i = \mathbf{X}_i\boldsymbol{\beta} + \boldsymbol{\mu}_\mathbf{X}\boldsymbol{\lambda}_i + \mathbf{e}_i \tag{3.1}$$

$$\boldsymbol{\mu}_\mathbf{X} = E(\mathbf{X}_i), \text{rank}(\boldsymbol{\mu}_\mathbf{X}) = k \tag{3.2}$$

$$\boldsymbol{\lambda}_i \equiv \boldsymbol{\Lambda}_\mathbf{X}\boldsymbol{\gamma}_i \tag{3.3}$$

In order to allow $\boldsymbol{\lambda}_i$ ($\boldsymbol{\gamma}_i$) to be arbitrarily correlated with $\mathbf{X}_i$, we would like to sweep away the term $\boldsymbol{\mu}_\mathbf{X}\boldsymbol{\lambda}_i$. If we knew $\boldsymbol{\mu}_\mathbf{X}$, we could pre-multiply (3.1) by the residual making matrix $\mathbf{M}_{\boldsymbol{\mu}_\mathbf{X}} = \mathbf{I}_T - \boldsymbol{\mu}_\mathbf{X}(\boldsymbol{\mu}_\mathbf{X}'\boldsymbol{\mu}_\mathbf{X})^{-1}\boldsymbol{\mu}_\mathbf{X}'$. Instead, CCEP replaces $\boldsymbol{\mu}_\mathbf{X}$ with the cross-sectional sample averages,

$$\bar{\mathbf{X}} = N^{-1}\sum_{i=1}^{N}\mathbf{X}_i = \begin{pmatrix} \bar{\mathbf{x}}_1 \\ \bar{\mathbf{x}}_2 \\ \vdots \\ \bar{\mathbf{x}}_T \end{pmatrix}. \tag{3.4}$$

By the law of large numbers

$$\bar{\mathbf{X}} \xrightarrow{p} \boldsymbol{\mu}_\mathbf{X} \text{ as } N \to \infty; \tag{3.5}$$

given the rank condition (3.2), $\bar{\mathbf{X}}'\bar{\mathbf{X}}$ is nonsingular with probability approaching one (WPA1). Therefore, define (WPA1)

$$\mathbf{M}_{\bar{\mathbf{X}}} = \mathbf{I}_T - \bar{\mathbf{X}}(\bar{\mathbf{X}}'\bar{\mathbf{X}})^{-1}\bar{\mathbf{X}}', \tag{3.6}$$

the residual-making matrix from regressing onto $\bar{\mathbf{X}}$. The resulting CCEP estimator can be written as

$$\hat{\boldsymbol{\beta}}_{CCEP(\bar{\mathbf{X}})} = \left(\sum_{i=1}^{N}\mathbf{X}_i'\mathbf{M}_{\bar{\mathbf{X}}}\mathbf{X}_i\right)^{-1}\left(\sum_{i=1}^{N}\mathbf{X}_i'\mathbf{M}_{\bar{\mathbf{X}}}\mathbf{y}_i\right), \tag{3.7}$$

where $CCEP(\bar{\mathbf{X}})$ indicates we sweep away the elements of $\bar{\mathbf{X}}$ in the unit-specific regressions. Technically, we should define $\hat{\boldsymbol{\beta}}_{CCEP(\bar{\mathbf{X}})}$ when $\bar{\mathbf{X}}'\bar{\mathbf{X}}$ is singular, but our definition has no impact on the asymptotic analysis. A necessary condition for (3.7) to exist is $T > k$; otherwise $\mathbf{M}_{\bar{\mathbf{X}}}\mathbf{X}_i = \mathbf{0}$.

Mechanically, $\hat{\boldsymbol{\beta}}_{CCEP(\bar{\mathbf{X}})}$ can be obtained as follows. First, for each unit $i$, run the multivariate regression

$$\mathbf{x}_{it} \text{ on } \bar{\mathbf{x}}_t, \ t = 1,\ldots,T \qquad (3.8)$$

and obtain the $1 \times k$ residuals, $\ddot{\mathbf{x}}_{it}$. Second, obtain $\hat{\boldsymbol{\beta}}_{CCEP(\bar{\mathbf{X}})}$ as the pooled OLS estimator from the regression

$$y_{it} \text{ on } \ddot{\mathbf{x}}_{it}, \ t = 1,\ldots,T; \ i = 1,\ldots,N. \qquad (3.9)$$

The estimator is unchanged if we also use the unit-specific regressions

$$y_{it} \text{ on } \bar{\mathbf{x}}_t, \ t = 1,\ldots,T$$

and obtain residuals $\ddot{y}_{it}$ and use these in place of $y_{it}$ in (3.9).

To further understand the nature of the CCEP estimator, note that if in (3.7) we replace $\bar{\mathbf{x}}_t$ with unity then the POLS regression (3.8) produces the usual within (fixed effects) estimator. If we use $(1, t)$ in place of $\bar{\mathbf{x}}_t$ then we obtain an estimator that removes unit-specific linear time trends, as in Wooldridge (2005).

To obtain additional assumptions needed for consistency, plug (3.1) into (3.7) and use simple algebra to write

$$\hat{\boldsymbol{\beta}}_{CCEP(\bar{\mathbf{X}})} = \boldsymbol{\beta} + \left( N^{-1} \sum_{i=1}^{N} \mathbf{X}_i' \mathbf{M}_{\bar{\mathbf{X}}} \mathbf{X}_i \right)^{-1} \left[ \left( N^{-1} \sum_{i=1}^{N} \mathbf{X}_i' \mathbf{M}_{\bar{\mathbf{X}}} \boldsymbol{\mu}_{\mathbf{X}} \boldsymbol{\lambda}_i \right) + \left( N^{-1} \sum_{i=1}^{N} \mathbf{X}_i' \mathbf{M}_{\bar{\mathbf{X}}} \mathbf{e}_i \right) \right]. \qquad (3.10)$$

By (3.5) and Slutsky's Theorem, $\mathbf{M}_{\bar{\mathbf{X}}} \xrightarrow{p} \mathbf{M}_{\boldsymbol{\mu}_{\mathbf{X}}}$. By standard arguments, which we formulate more generally in Section 5, for consistency we can effectively replace the random matrix $\mathbf{M}_{\bar{\mathbf{X}}}$ with its nonrandom plim, $\mathbf{M}_{\boldsymbol{\mu}_{\mathbf{X}}}$. Then we can apply the LLN to each of the three averages:

$$N^{-1} \sum_{i=1}^{N} \mathbf{X}'_i \mathbf{M}_{\bar{\mathbf{X}}} \mathbf{X}_i \xrightarrow{p} E(\mathbf{X}'_i \mathbf{M}_{\mu_\mathbf{X}} \mathbf{X}_i)$$

$$N^{-1} \sum_{i=1}^{N} \mathbf{X}'_i \mathbf{M}_{\bar{\mathbf{X}}} \mu_\mathbf{X} \lambda_i \xrightarrow{p} E(\mathbf{X}'_i \mathbf{M}_{\mu_\mathbf{X}} \mu_\mathbf{X} \lambda_i)$$

$$N^{-1} \sum_{i=1}^{N} \mathbf{X}'_i \mathbf{M}_{\bar{\mathbf{X}}} \mathbf{e}_i \xrightarrow{p} E(\mathbf{X}'_i \mathbf{M}_{\mu_\mathbf{X}} \mathbf{e}_i)$$

The second expectation is zero because $\mathbf{M}_{\mu_\mathbf{X}} \mu_\mathbf{X} = \mathbf{0}$. Therefore, for consistency of $\hat{\boldsymbol{\beta}}_{CCEP}$ for fixed $T$, $N \to \infty$, it suffices to assume $E(\mathbf{X}'_i \mathbf{M}_{\mu_\mathbf{X}} \mathbf{X}_i)$ is nonsingular and $E(\mathbf{X}'_i \mathbf{M}_{\mu_\mathbf{X}} \mathbf{e}_i) = \mathbf{0}$. The former is a standard rank condition, which requires sufficient variation in $\{\mathbf{x}_{it} : t = 1, \ldots, T\}$ after partialling out $\mu_\mathbf{X}$; we will have more to say about it in Section 5. The latter assumption can be written as

$$E(\dot{\mathbf{X}}'_i \mathbf{e}_i) = E(\dot{\mathbf{X}}'_i \dot{\mathbf{e}}_i) = \mathbf{0} \tag{3.11}$$

where $\dot{\mathbf{X}}_i \equiv \mathbf{M}_{\mu_\mathbf{X}} \mathbf{X}_i$ and $\dot{\mathbf{e}}_i \equiv \mathbf{M}_{\mu_\mathbf{X}} \mathbf{e}_i$. A sufficient but not necessary condition for (3.11) is that $\{\mathbf{x}_{it} : t = 1, \ldots, T\}$ is strictly exogenous with respect to $\{e_{it} : t = 1, \ldots, T\}$, a standard assumption in the factor model literature in general, including CCE.

The asymptotic distribution of $\sqrt{N}\left(\hat{\boldsymbol{\beta}}_{CCEP(\bar{\mathbf{X}})} - \boldsymbol{\beta}\right)$ is complicated by the fact that it generally depends on the asymptotic distribution of $\sqrt{N}(\mathbf{M}_{\bar{\mathbf{X}}} - \mathbf{M}_{\mu_\mathbf{X}})$, something that has been thus far ignored in the CCE literature. Ignoring the sampling error in $\bar{\mathbf{X}}$ generally leads to incorrect calculation of standard errors and test statistics. In Sections 5 and 6 we will show consistency of an estimator in more general settings and obtain its $\sqrt{N}$-asymptotically normal distribution in Section 7.

## 3.2. Adding $\bar{y}$ in CCEP

Pesaran (2006) proposed a version of CCEP that includes the cross-sectional averages of the outcome variable, and this estimator has been studied in the small-$T$ literature under SCF. By extending the previous analysis we can derive its consistency under much weaker assumptions than those imposed in WPN (2019). Define

$$\bar{\mathbf{y}} = N^{-1} \sum_{i=1}^{N} \mathbf{y}_i = \begin{pmatrix} \bar{y}_1 \\ \bar{y}_2 \\ \vdots \\ \bar{y}_T \end{pmatrix} \tag{3.12}$$

and

$$\bar{\mathbf{Z}} = (\bar{\mathbf{y}}, \bar{\mathbf{X}}). \tag{3.13}$$

Assume that the $T \times (k+1)$ matrix $\boldsymbol{\mu}_{\mathbf{Z}} = (\boldsymbol{\mu}_{\mathbf{y}}, \boldsymbol{\mu}_{\mathbf{X}})$ has rank $k+1$ and that $\mathrm{E}(\mathbf{X}_i' \mathbf{M}_{\boldsymbol{\mu}_{\mathbf{Z}}} \mathbf{X}_i)$ is nonsingular; necessary is $T > k + 1$. The new estimator is

$$\hat{\boldsymbol{\beta}}_{CCEP(\bar{\mathbf{Z}})} = \left( \sum_{i=1}^{N} \mathbf{X}_i' \mathbf{M}_{\bar{\mathbf{Z}}} \mathbf{X}_i \right)^{-1} \left( \sum_{i=1}^{N} \mathbf{X}_i' \mathbf{M}_{\bar{\mathbf{Z}}} \mathbf{y}_i \right) \tag{3.14}$$

Consistency of this estimator follows from essentially the same argument in Section 3.

The CCEP estimator with $\bar{\mathbf{y}}$ included imposes a cost in terms of a lost degree of freedom. Nevertheless, there is a situation, covered in WPN (2019) in the fixed $T$ case, that warrants inclusion of $\bar{\mathbf{y}}$ in approximating the factors. This is the case $p = k + 1$ in the Pesaran (2006) and WPN (2019) frameworks, so that there is one more factor than regressor. Under the WPN (2019) assumptions, rank$(\mathbf{F}) = k + 1$. (We will discuss the rank condition on $\mathbf{F}$ further in Section 9.) Given (2.2), $\mathbf{y}_i = \mathbf{X}_i \boldsymbol{\beta} + \mathbf{F} \boldsymbol{\gamma}_i + \mathbf{e}_i$, $\mathrm{E}(\mathbf{e}_i) = \mathbf{0}$, and the WCF assumption (2.6), we have

$$\mu_y = \mu_X \beta + F\gamma \tag{3.15}$$

$$\mu_X = F\Gamma \tag{3.16}$$

where $\gamma \equiv E(\gamma_i)$. [WPN (2019) actually impose SCF, but the analysis here shows those strong assumptions are not needed.] This implies we can write

$$(\mu_y, \mu_X) = FCQ \tag{3.17}$$

where

$$C \equiv (\gamma, \Gamma) \tag{3.18}$$

$$Q \equiv \begin{pmatrix} 1 & 0 \\ \beta & I_k \end{pmatrix} \tag{3.19}$$

Letting $\mu_Z \equiv (\mu_y, \mu_X)$, with $t^{th}$ row $\mu_t^z \equiv (\mu_t^y, \mu_t^x)$, and $\Pi \equiv CQ$, we can write

$$\mu_Z = F\Pi \tag{3.20}$$

The matrix $Q$ is always nonsingular. Further, nonsingularity of $C$ is assumed in WPN (2019) and it is a natural assumption. Therefore, we can write

$$F = \mu_Z \Pi^{-1}$$

or

$$F = \mu_Z \Lambda_Z \tag{3.21}$$

The representation in (3.21) is the natural counterpart to the earlier BSW assumption (2.7). Of course (3.21) nests (2.7), but that does not mean we can or want to include $\bar{y}$ in the CCEP estimation. We face a tradeoff between having more estimable proxies for the factors $f_t$ and reducing the remaining variation in $\{x_{it} : t = 1, \ldots, T\}$. And, including $\bar{y}$ is not possible if $T = k+1$.

We have shown that under (2.2) with $p = k + 1$ and $\mathbf{C}$ nonsingular, (3.20) and (3.21) are equivalent. Because these assumptions imply $\text{rank}(\boldsymbol{\mu}_\mathbf{Z}) = k + 1$, $\hat{\boldsymbol{\beta}}_{CCEP(\bar{\mathbf{Z}})}$ is defined WPA1 and it is consistent for $\boldsymbol{\beta}$; we need not worry about difficulties caused by degeneracies.

## 4. Adding Observed Deterministic Variables

A natural extension of model (2.2) is to add a (row) vector of known variables $\mathbf{d}_t$ $(1 \times r)$, which can include an overall intercept, time period dummies, and other variables that change only across $t$. Then (3.1) gets replaced with

$$\mathbf{y}_i = \mathbf{D}\boldsymbol{\alpha} + \mathbf{X}_i\boldsymbol{\beta} + \boldsymbol{\mu}_\mathbf{X}\boldsymbol{\lambda}_i + \mathbf{e}_i \tag{4.1}$$

where $\mathbf{D}$ is a $T \times r$ matrix with $t^{th}$ row $\mathbf{d}_t$. When $\bar{\mathbf{y}}$ is included in CCEP the relevant model, reusing notation, is

$$\mathbf{y}_i = \mathbf{D}\boldsymbol{\alpha} + \mathbf{X}_i\boldsymbol{\beta} + \boldsymbol{\mu}_\mathbf{Z}\boldsymbol{\lambda}_i + \mathbf{e}_i \tag{4.2}$$

However, whether any particular model holds is irrelevant as the following result is purely algebraic. In what follows, $\ddot{\mathbf{X}}$ is the $NT \times k$ matrix obtained by stacking the $\ddot{\mathbf{X}}_i = \mathbf{M}_{\bar{\mathbf{X}}} \mathbf{X}_i$ over $i = 1, \ldots, N$.

**THEOREM 1**: Assume that $\ddot{\mathbf{X}}$ has rank $k$. Further, assume that the $T \times r$ matrix $\ddot{\mathbf{D}} = \mathbf{M}_{\bar{\mathbf{X}}} \mathbf{D}$ has rank $r$.

(i) Let $\hat{\boldsymbol{\alpha}}$ and $\hat{\boldsymbol{\beta}}$ be the estimators obtained by partialling out $\bar{\mathbf{X}}$ at the unit level. Then $\hat{\boldsymbol{\beta}} = \hat{\boldsymbol{\beta}}_{CCEP(\bar{\mathbf{X}})}$.

(ii) If $\hat{\boldsymbol{\alpha}}$ and $\hat{\boldsymbol{\beta}}$ are obtained by partialling out $\bar{\mathbf{Z}}$ at the unit level then $\hat{\boldsymbol{\beta}} = \hat{\boldsymbol{\beta}}_{CCEP(\bar{\mathbf{Z}})}$ and $\hat{\boldsymbol{\alpha}} = \mathbf{0}$.

**Proof**: (i) By Frisch-Waugh partialling out, the estimates $\hat{\boldsymbol{\alpha}}$ and $\hat{\boldsymbol{\beta}}$ are obtained from the system OLS regression

$$\ddot{\mathbf{y}}_i \text{ on } \ddot{\mathbf{D}}, \ddot{\mathbf{X}}_i, i = 1,\ldots,N, \tag{4.3}$$

where $\ddot{\mathbf{D}} = \mathbf{M}_{\bar{\mathbf{X}}}\mathbf{D}$ and $\ddot{\mathbf{X}}_i = \mathbf{M}_{\bar{\mathbf{X}}}\mathbf{X}_i$. By the rank conditions and the algebra of partitioned least squares, this long regression gives the same coefficients on $\ddot{\mathbf{X}}_i$ in the short (CCEP) regression if $\sum_{i=1}^{N} \ddot{\mathbf{D}}' \ddot{\mathbf{X}}_i = \mathbf{0}$. But

$$\sum_{i=1}^{N} \ddot{\mathbf{D}}' \ddot{\mathbf{X}}_i = \ddot{\mathbf{D}}' \sum_{i=1}^{N} \ddot{\mathbf{X}}_i = \ddot{\mathbf{D}}' \sum_{i=1}^{N} \mathbf{M}_{\bar{\mathbf{X}}} \mathbf{X}_i = \ddot{\mathbf{D}}'(N\mathbf{M}_{\bar{\mathbf{X}}}\bar{\mathbf{X}}) = \mathbf{0}. \tag{4.4}$$

The same proof shows that it is irrelevant whether we partial $\bar{\mathbf{X}}$ out of $\mathbf{D}$: the estimate $\hat{\boldsymbol{\beta}}_{CCEP(\bar{\mathbf{X}})}$ is also obtained from $\ddot{\mathbf{y}}_i$ on $\mathbf{D}, \ddot{\mathbf{X}}_i, i = 1,\ldots,N$.

(ii) $\hat{\boldsymbol{\alpha}}$ and $\hat{\boldsymbol{\beta}}$ are obtained from (4.3) but where $\ddot{\mathbf{D}} = \mathbf{M}_{\bar{\mathbf{Z}}}\mathbf{D}$, $\ddot{\mathbf{X}}_i = \mathbf{M}_{\bar{\mathbf{Z}}}\mathbf{X}_i$, and $\ddot{\mathbf{y}}_i = \mathbf{M}_{\bar{\mathbf{Z}}}\mathbf{y}_i$. Because the columns of $\bar{\mathbf{Z}}$ include $\bar{\mathbf{X}}$, $\mathbf{M}_{\bar{\mathbf{Z}}}\bar{\mathbf{X}} = \mathbf{0}$ and so (4.4) still holds, which implies $\hat{\boldsymbol{\beta}} = \hat{\boldsymbol{\beta}}_{CCEP(\bar{\mathbf{Z}})}$ and

$$\hat{\boldsymbol{\alpha}} = \left(\sum_{i=1}^{N} \ddot{\mathbf{D}}' \ddot{\mathbf{D}}\right)^{-1} \left(\sum_{i=1}^{N} \ddot{\mathbf{D}}' \ddot{\mathbf{y}}_i\right)$$

But now, using $\mathbf{M}_{\bar{\mathbf{Z}}}\bar{\mathbf{y}} = \mathbf{0}$,

$$\sum_{i=1}^{N} \ddot{\mathbf{D}}' \ddot{\mathbf{y}}_i = \ddot{\mathbf{D}}' \sum_{i=1}^{N} \ddot{\mathbf{y}}_i = \ddot{\mathbf{D}}' \sum_{i=1}^{N} \mathbf{M}_{\bar{\mathbf{Z}}}\mathbf{y}_i = \ddot{\mathbf{D}}'(N\mathbf{M}_{\bar{\mathbf{Z}}}\bar{\mathbf{y}}) = \mathbf{0}$$

which implies $\hat{\boldsymbol{\alpha}} = \mathbf{0}$. $\square$

Theorem 1 shows that once the $\bar{\mathbf{x}}_t$ have been netted out of the $\mathbf{x}_{it}$ no other aggregate factors matter for estimation of $\boldsymbol{\beta}$. One interpretation is that these are the only relevant factors for estimating the effects of the $\mathbf{x}_{it}$. Commonly used estimators in small-$T$ settings, such as the standard fixed effects (or within) estimator, are not invariant to the inclusion of $\mathbf{D}$. In fact, there are often notable differences between the FE estimator that excludes time effects and the FE estimator that includes a full set of time dummies – the latter of which is often called the "two-way fixed effects" (TWFE) estimator. Part (ii) shows that if $\bar{\mathbf{y}}$ is included in CCEP then we can say more: the coefficients on any regressors $\mathbf{d}_t$ are identically zero. Consequently, when the unit-specific partialling out is done on $\bar{\mathbf{Z}}$, the residuals obtained from (4.3) are identical to those obtained by dropping $\ddot{\mathbf{D}}$.

## 5. Consistency of an Extended CCEP Estimator under a General Factor Structure

In Section 3 we sketched the consistency of the versions of CCEP that use $\bar{\mathbf{X}}$ or $(\bar{\mathbf{y}}, \bar{\mathbf{X}})$ under fixed-$T$ asymptotics under conditions substantially weaker than WPN (2019). For $\hat{\boldsymbol{\beta}}_{CCEP(\bar{\mathbf{X}})}$, we simply require that $\boldsymbol{\mu}_\mathbf{X}$ has rank $k$. Likewise, for $\hat{\boldsymbol{\beta}}_{CCEP(\bar{\mathbf{Z}})}$ we only require that $(\boldsymbol{\mu}_y, \boldsymbol{\mu}_\mathbf{X})$ has rank $k + 1$. In Section 4 we showed that, in either case, estimation of $\boldsymbol{\beta}$ is invariant to the inclusion of deterministic variables $\mathbf{d}_t$ in the equation of interest.

There is still a shortcoming of the CCEP estimators in Sections 3 and 4 for microeconometric application: they rule out staples of empirical applications of small-$T$ panel data analysis, such as unit-specific intercepts. Some microeconometric applications allow for unit-specific time trends, too. Therefore, as in Pesaran (2006) and WPN (2019), in this section we study consistency of a class of estimators that includes both versions of the CCEP estimators as well as extensions of CCEP that allow for additive heterogeneity and heterogenous trends (linear or otherwise). These can be viewed as hybrids between standard FE-type estimators and CCEP.

We also extend the framework to allow for other estimable quantities to appear in the factor structure. For example, with enough time periods we can allow squares and interactions among elements in $(\boldsymbol{\mu}_\mathbf{X}, \boldsymbol{\mu}_y)$, something that has not been previously allowed (probably because both SCF and WCF rule out this possibility). Moreover, we can include cross-sectional second moments of $\{(\mathbf{x}_{it}, y_{it}) : i = 1, \ldots, N\}$ as additional proxies for the factors.

In Section 4 we saw that inclusion of variables that change only across $t$ does not change either of the CCEP estimates. This is not the same as saying the estimators that omit these variables are consistent for $\boldsymbol{\beta}$ in a general model. Therefore, in this section we explicitly allow for a vector of variables $\mathbf{d}_t$ in the model and show that ignoring these factors results in consistent estimation.

The model is now assumed to be

$$y_{it} = \mathbf{d}_t \boldsymbol{\alpha} + \mathbf{x}_{it} \boldsymbol{\beta} + \mathbf{f}_t \boldsymbol{\gamma}_i + e_{it}, \, t = 1, \ldots, T \tag{5.1}$$

or, stacking over time,

$$\mathbf{y}_i = \mathbf{D}\boldsymbol{\alpha} + \mathbf{X}_i \boldsymbol{\beta} + \mathbf{F}\boldsymbol{\gamma}_i + \mathbf{e}_i \tag{5.2}$$

where $\mathbf{D}$ is $T \times r$ (and nonrandom), $\mathbf{X}_i$ is $T \times k$, and $\mathbf{F}$ is $T \times p$ (and nonrandom). We now assume the factors are determined by

$$\mathbf{F} = \boldsymbol{\Psi}\boldsymbol{\Lambda}, \tag{5.3}$$

where $\boldsymbol{\Psi}$ is $T \times m$ and $\boldsymbol{\Lambda}$ is $m \times p$ and includes at least $\boldsymbol{\mu}_\mathbf{X}$.

Combining (5.2) and (5.3) we can write

$$y_{it} = \mathbf{d}_t \boldsymbol{\alpha} + \mathbf{x}_{it} \boldsymbol{\beta} + \boldsymbol{\psi}_t \boldsymbol{\lambda}_i + e_{it}, \, t = 1, \ldots, T \tag{5.4}$$

or

$$\mathbf{y}_i = \mathbf{D}\boldsymbol{\alpha} + \mathbf{X}_i \boldsymbol{\beta} + \boldsymbol{\Psi}\boldsymbol{\lambda}_i + \mathbf{e}_i \tag{5.5}$$

where

$$\lambda_i \equiv \Lambda \gamma_i \tag{5.6}$$

As can be seen from (5.5), the matrix $\Psi$ is key: it contains either known constants or consistently estimable parameters that determine the unknown factors $\mathbf{F}$. We assume throughout that $\mu_X$ is included in $\Psi$, which may also include $\mu_y$. Moreover, $\Psi$ can include observed, deterministic functions (such as unity or functions of time). Because we choose $\Psi$ we are necessarily choosing $m$. The number of factors, $p$, is unknown and unimportant to our approach – a significant advantage over other approaches to factor models.

It may seem that one way of allowing heterogeneous slopes and trends in CCEP is to simply include unity and functions of $t$ in $\mathbf{x}_{it}$. But this will always cause degeneracies because those elements of $\mathbf{x}_{it}$ will be swept away in the partialling out used by CCEP. Rather than deal with deficient rank situations, we follow the large-$T$ approach in Pesaran (2006) and include in $\mathbf{x}_{it}$ only variables that have some variation across $i$ and $t$. We allow more flexibility than basic CCE framework in Sections 2 and 3 by choosing $\Psi$ appropriately.

A natural extension of the CCEP estimator that projects out the cross-sectional averages of the explanatory variables is to take

$$\psi_t = (1, \mu_t^x), \tag{5.7}$$

a possibility allowed in Pesaran (2006) in the large-$T$ setting under much stronger assumptions than we make here. In the fixed-$T$ setting, WPN (2019) mention such possibilities but do not provide a formal analysis. By choosing $\psi_t$ as in (5.7) we allow for a heterogenous intercept in (5.4), $\lambda_{i1}$ – in addition to allowing the factors to depend heterogeneously on $\mu_t^x$. We obtain a kind of hybrid between the usual additive unobserved effects model and a version of the CCE model (without imposing the strong common factor assumption). Choosing $\psi_t$ in (5.7) requires $T > k + 1 = \dim(\mathbf{x}_{it}) + 1$, whereas the CCEP estimator that projects onto $\mathbf{\mu_X}$ (actually, its sample analog, $\bar{\mathbf{X}}$) only requires $T > k$.

Adding unit-specific linear trends to CCEP($\bar{\mathbf{X}}$) means taking

$$\mathbf{\psi}_t = (1, t, \mathbf{\mu}_t^x), \tag{5.8}$$

and then we require $T > k + 2$. Naturally, there is a tradeoff between the amount of heterogeneity allowed and the required number of time periods. If we were to drop $\mathbf{\mu}_t^x$ then we obtain the heterogenous trend model studied in Wooldridge (2005). Here we only consider cases where $\mathbf{\mu}_t^x$ is included in $\mathbf{\psi}_t$. We can also add $\mu_t^y$ to $\mathbf{\psi}_t$ in (5.7) or (5.8) and then we require an additional time period. With large enough $T$, more flexibility is obtained by including, say, $\mathbf{\mu}_t^x \otimes \mathbf{\mu}_t^x$, $\mathbf{\mu}_t^x \otimes \mu_t^y$, or even second moments of $(\mathbf{x}_{it}, y_{it})$. If squares and interactions appear in $\mathbf{x}_{it}$ then cross-sectional second moments are already included in $\mathbf{\mu}_t^x$, which is itself an extension from the previous frameworks that impose SCF. By directly specifying (5.7), we are not limited in what we can include in $\mathbf{\psi}_t$ by the restrictive SCF or WCF assumptions; we are only limited by the number of time periods.

In what follows, we assume $\{(\mathbf{X}_i, \mathbf{y}_i) : i = 1, 2, \ldots, N\}$ are independent and identically distributed. As in Section 3, the asymptotic analysis is for fixed $T$ with $N \to \infty$, and so we do not need restrictions on the distributions across time or the amount of time dependence. In order to focus on substantive assumptions we are not explicit about moment conditions that ensure the basic convergence results. These conditions are standard and add nothing to the analysis.

**Assumption 1 (Factor Model):** For a random draw $i$, the model is given by equation (5.2) where $\mathbf{D}$ ($T \times r$) and $\mathbf{F}$ ($T \times p$) are nonrandom, as are the parameters $\mathbf{\alpha}$ ($r \times 1$) and $\mathbf{\beta}$ ($k \times 1$). $\square$

Importantly, we put no restrictions on how many elements of $\gamma_i$ are correlated with elements of $\mathbf{X}_i$. It could be all elements of $\gamma_i$ or none. In other approaches to estimating $\beta$ in (5.2) the the number of factors is defined to be number of heterogeneity terms correlated with $\mathbf{X}_i$, and this integer value needs to be known or estimated. See, for example, Ahn, Lee, and Schmidt (2013).

**Assumption 2 (Factor Restriction)**: The $T \times p$ matrix of factors $\mathbf{F}$ is given by $\mathbf{F} = \Psi \Lambda$ for some $T \times m$ matrix $\Psi$ with $\mu_X \in \Psi$ and $\text{rank}(\Psi) = m$. $\square$

As discussed earlier, a leading case of Assumption 2 is when $\Psi = \mu_X$. Assumption 2 allows one to include $\mu_y$ in $\Psi$ as well as a constant and time trends. One can even include other cross sectional moments, such as standard deviations of the elements of $\mathbf{x}_{it}$. Because $\Psi$ at least includes $\mu_X$ it is necessarily true that $m \geq k$. Importantly, once we have chosen $\Psi$ the column dimension of $\mathbf{F}$, $p$, plays no role, and neither does the matrix $\Lambda$. This is in contrast to other fixed-$T$ approaches, notably WPN (2019). The assumption $\text{rank}(\Psi) = m$ seems reasonable to us and makes the asymptotic analysis relatively straightforward.

Given the algebraic equivalences in Theorem 1, if interest lies in only estimating $\beta$ then we could ignore the presence of $\mathbf{D}\alpha$. Nevertheless, in some cases it is useful to know that $\alpha$ is consistently estimated, especially when one wants to obtain suitable residuals.

**Assumption 3 (Rank Condition)**: Define

$$\mathbf{M}_\Psi \equiv \mathbf{I}_T - \Psi(\Psi'\Psi)^{-1}\Psi'. \tag{5.9}$$

Then (i) $\text{E}\left(\dot{\mathbf{X}}_i'\dot{\mathbf{X}}_i\right) \equiv \text{E}(\mathbf{X}_i'\mathbf{M}_\Psi \mathbf{X}_i)$ is nonsingular and (ii) $\dot{\mathbf{D}}'\dot{\mathbf{D}} = \mathbf{D}'\mathbf{M}_\Psi \mathbf{D}$ is nonsingular. $\square$

The first part of Assumption 3 serves to identify $\beta$ and the second part is needed only to identify $\alpha$. Combined with Assumption 2, Assumption 3(i) requires $T > m = \dim(\psi_t)$ – which we can think of as an order condition – so that projecting $\mathbf{X}_i$ onto $\Psi$ (in the population) does not result in a perfect fit. The order condition is a natural restriction as it simply says that the number of (estimable) quantities determining the factors does not exceed the number of time periods. Moreover, this condition is entirely consistent with what is required in simpler fixed effects and heterogeneous trend settings. In order to satisfy Assumption 3(ii), the choice of $\mathbf{D}$ is restricted by our choice of $\Psi$ because $\text{rank}(\ddot{\mathbf{D}}) \leq \text{rank}(\mathbf{M}_\Psi) = T - m$. Therefore, there is no point in putting more than $T - m$ elements in $\mathbf{D}$. Typically, each row $\mathbf{d}_t$ would include time period dummy variables, but some time period dummies will be redundant. Remember, whatever we put in $\mathbf{D}$ does not affect estimation of $\boldsymbol{\beta}$. Generally, it will be difficult to interpret the $\hat{\boldsymbol{\alpha}}$ when it is not identically zero.

As discussed previously, useful extensions of the basic CCEP estimator are obtained by choosing

$$\boldsymbol{\psi}_t = (\mathbf{g}_t, \boldsymbol{\mu}_t^x), \tag{5.10}$$

where $\mathbf{g}_t$ includes known factors, such as unity (the leading case) and functions of time. No matter what is in $\mathbf{g}_t$, we still need $T > \dim(\boldsymbol{\psi}_t) = m$; otherwise $\mathbf{M}_\Psi \mathbf{X}_i = \mathbf{0}$.

If we know $\Psi$ then we can premultiply (5.5) by $\mathbf{M}_\Psi$ to remove $\boldsymbol{\lambda}_i$:

$$\mathbf{M}_\Psi \mathbf{y}_i = \mathbf{M}_\Psi \mathbf{D} \boldsymbol{\alpha} + \mathbf{M}_\Psi \mathbf{X}_i \boldsymbol{\beta} + \mathbf{M}_\Psi \mathbf{e}_i$$

or

$$\dot{\mathbf{y}}_i = \dot{\mathbf{D}} \boldsymbol{\alpha} + \dot{\mathbf{X}}_i \boldsymbol{\beta} + \dot{\mathbf{e}}_i \tag{5.11}$$

where the "·" denotes residuals from regressing onto $\Psi$. The extended CCEP estimator with known $\Psi$ is the system OLS estimator applied to (5.11).

**Assumption 4 (Exogeneity):** With $\dot{\mathbf{X}}_i \equiv \mathbf{M}_\Psi \mathbf{X}_i$,

$$E(\dot{\mathbf{e}}_i) = \mathbf{0} \tag{5.12}$$

$$E(\dot{\mathbf{X}}_i' \mathbf{e}_i) = \mathbf{0}. \ \square \tag{5.13}$$

Because $\mathbf{D}$ can include an intercept and time dummies the assumption $E(\dot{\mathbf{e}}_i) = \mathbf{0}$ is for free. In any case, it only has implications for estimating $\boldsymbol{\alpha}$. To see why, if we define $\boldsymbol{\mu}_\mathbf{e} \equiv E(\mathbf{e}_i)$ then

$$E\left[\dot{\mathbf{X}}_i'(\mathbf{e}_i - \boldsymbol{\mu}_\mathbf{e})\right] = E(\dot{\mathbf{X}}_i' \mathbf{e}_i) - E(\dot{\mathbf{X}}_i')\boldsymbol{\mu}_\mathbf{e} = E(\dot{\mathbf{X}}_i' \mathbf{e}_i) - \boldsymbol{\mu}_\mathbf{X}' \mathbf{M}_\Psi \boldsymbol{\mu}_\mathbf{e} = E(\dot{\mathbf{X}}_i' \mathbf{e}_i)$$

because $\boldsymbol{\mu}_\mathbf{X}' \mathbf{M}_\Psi = \mathbf{0}$ when $\boldsymbol{\mu}_\mathbf{X} \in \Psi$. Therefore, the exogeneity assumption (5.13), which is the one relevant for identification of $\boldsymbol{\beta}$, holds without (5.12).

Condition (5.13) is implied by the strict exogeneity assumption

$$\text{Cov}(\mathbf{x}_{is}, e_{it}) = \mathbf{0}, \ s, t = 1, \ldots, T, \tag{5.14}$$

which is a common assumption in the CCE literature.

As a preamble to establishing consistency of a feasible estimator, it is useful to begin by assuming $\Psi$ is known. Then, it is useful to write

$$\mathbf{y}_i = \mathbf{W}_i \boldsymbol{\theta} + \Psi \boldsymbol{\lambda}_i + \mathbf{e}_i \tag{5.15}$$

where

$$\mathbf{W}_i \equiv (\mathbf{D}, \mathbf{X}_i), \ \boldsymbol{\theta} \equiv (\boldsymbol{\alpha}', \boldsymbol{\beta}')'.$$

The system OLS estimator from (5.11) can be written as

$$\hat{\boldsymbol{\theta}} = \left(\sum_{i=1}^N \dot{\mathbf{W}}_i' \dot{\mathbf{W}}_i\right)^{-1} \left(\sum_{i=1}^N \dot{\mathbf{W}}_i' \dot{\mathbf{y}}_i\right) = \left(\sum_{i=1}^N \mathbf{W}_i' \mathbf{M}_\Psi \mathbf{W}_i\right)^{-1} \left(\sum_{i=1}^N \mathbf{W}_i' \mathbf{M}_\Psi \mathbf{y}_i\right)$$

$$= \boldsymbol{\theta} + \left(N^{-1} \sum_{i=1}^N \mathbf{W}_i' \mathbf{M}_\Psi \mathbf{W}_i\right)^{-1} \left(N^{-1} \sum_{i=1}^N \mathbf{W}_i' \mathbf{M}_\Psi \mathbf{e}_i\right)$$

where we use $\mathbf{M}_\Psi \Psi = \mathbf{0}$. By the law of large numbers and $E(\dot{\mathbf{D}}' \dot{\mathbf{X}}_i) = \dot{\mathbf{D}}' E(\dot{\mathbf{X}}_i) = \mathbf{0}$,

$$N^{-1} \sum_{i=1}^{N} \mathbf{W}_i' \mathbf{M}_\Psi \mathbf{W}_i \overset{p}{\to} E(\mathbf{W}_i' \mathbf{M}_\Psi \mathbf{W}_i) = \begin{pmatrix} \dot{\mathbf{D}}'\dot{\mathbf{D}} & 0 \\ 0 & E(\dot{\mathbf{X}}_i'\dot{\mathbf{X}}_i) \end{pmatrix},$$

which is nonsingular by Assumption 3. (This shows that $\hat{\boldsymbol{\theta}}$ exists WPA1.) Next, by the LLN,

$$N^{-1} \sum_{i=1}^{N} \mathbf{W}_i' \mathbf{M}_\Psi \mathbf{e}_i = N^{-1} \sum_{i=1}^{N} \dot{\mathbf{W}}_i' \mathbf{e}_i \overset{p}{\to} \begin{pmatrix} \dot{\mathbf{D}}' E(\mathbf{e}_i) \\ E(\dot{\mathbf{X}}_i' \mathbf{e}_i) \end{pmatrix} = \mathbf{0}$$

by Assumption 4. We have shown that, with known $\Psi$, the CCEP estimator is consistent for $\alpha$ and $\beta$ with fixed $T$, $N \to \infty$ under Assumptions 1 through 4.

We now turn to a formal consistency result for a feasible estimator that replaces unknown elements in $\Psi$ with consistent estimators.

**Assumption 5 (Consistent Estimation of $\Psi$):** For an estimator $\hat{\Psi}$,

$$\hat{\Psi} \overset{p}{\to} \Psi \text{ as } N \to \infty. \;\square \tag{5.16}$$

Because $\boldsymbol{\mu}_\mathbf{X} \in \Psi$, $\bar{\mathbf{X}}$ would be included in $\hat{\Psi}$. Therefore, by the same reasoning as at the end of Section 3, the feasible estimator is invariant to the inclusion of $\mathbf{D}$. Nevertheless, we show consistency of the estimator of $\alpha$ along with that of $\beta$. When $\hat{\Psi} = (\bar{\mathbf{X}}, \bar{\mathbf{y}})$ we obtain the version of the CCEP estimator that includes $\bar{\mathbf{y}}$. If $\boldsymbol{\psi}_t$ is chosen as in (5.10), $\hat{\Psi} = (\mathbf{G}, \bar{\mathbf{X}})$, where $\mathbf{G}$ is a $T \times q$ matrix of deterministic functions of $t$. As mentioned previously, other choices are possible, such as including cross-sectional variances and covariances of the elements of $\mathbf{x}_{it}$, although this would require a fairly large $T$ in most cases.

Intuitively, it is clear that, under Assumptions 1 through 5, replacing $\mathbf{M}_\Psi$ with

$$\mathbf{M}_{\hat{\Psi}} = \mathbf{I}_T - \hat{\Psi}(\hat{\Psi}'\hat{\Psi})^{-1}\hat{\Psi}' \tag{5.17}$$

will not affect consistency of the estimator, which is now written

$$\hat{\boldsymbol{\theta}} = \left(\sum_{i=1}^{N} \mathbf{W}_i' \mathbf{M}_{\hat{\boldsymbol{\Psi}}} \mathbf{W}_i\right)^{-1} \left(\sum_{i=1}^{N} \mathbf{W}_i' \mathbf{M}_{\hat{\boldsymbol{\Psi}}} \mathbf{y}_i\right) \equiv \left(\sum_{i=1}^{N} \ddot{\mathbf{W}}_i' \ddot{\mathbf{W}}_i\right)^{-1} \left(\sum_{i=1}^{N} \ddot{\mathbf{W}}_i' \mathbf{y}_i\right) \quad (5.18)$$

$$\ddot{\mathbf{W}}_i \equiv \mathbf{M}_{\hat{\boldsymbol{\Psi}}} \mathbf{X}_i = (\ddot{\mathbf{D}} | \ddot{\mathbf{X}}_i). \quad (5.19)$$

Nevertheless, we state a formal result that covers many cases of interest, including any of the CCEP estimators.

**THEOREM 2**: Under Assumptions 1 through 5 and standard moment conditions, $\hat{\boldsymbol{\theta}} \xrightarrow{p} \boldsymbol{\theta}$ as $N \to \infty$. □

**Proof**: By Assumptions 2 and 5, $\mathbf{M}_{\hat{\boldsymbol{\Psi}}}$ exists WPA1. Moreover, by Slutsky's Theorem, $\mathbf{M}_{\hat{\boldsymbol{\Psi}}} \xrightarrow{p} \mathbf{M}_{\boldsymbol{\Psi}}$. It follows that

$$\text{vec}\left(N^{-1} \sum_{i=1}^{N} \mathbf{W}_i' \mathbf{M}_{\hat{\boldsymbol{\Psi}}} \mathbf{W}_i - N^{-1} \sum_{i=1}^{N} \mathbf{W}_i' \mathbf{M}_{\boldsymbol{\Psi}} \mathbf{W}_i\right) = \left[N^{-1} \sum_{i=1}^{N} (\mathbf{W}_i' \otimes \mathbf{W}_i')\right] \text{vec}(\mathbf{M}_{\hat{\boldsymbol{\Psi}}} - \mathbf{M}_{\boldsymbol{\Psi}})$$
$$= O_p(1) o_p(1) = o_p(1)$$

because $N^{-1} \sum_{i=1}^{N} (\mathbf{W}_i' \otimes \mathbf{W}_i')$ converges in probability to $E(\mathbf{W}_i' \otimes \mathbf{W}_i')$. It follows by the law of large numbers applied to $N^{-1} \sum_{i=1}^{N} \mathbf{W}_i' \mathbf{M}_{\boldsymbol{\Psi}} \mathbf{W}_i$ that

$$N^{-1} \sum_{i=1}^{N} \mathbf{W}_i' \mathbf{M}_{\hat{\boldsymbol{\Psi}}} \mathbf{W}_i \xrightarrow{p} E(\mathbf{W}_i' \mathbf{M}_{\boldsymbol{\Psi}} \mathbf{W}_i), \quad (5.20)$$

which is nonsingular by Assumption 3.

Next, plug in for $\mathbf{y}_i$:

$$\hat{\boldsymbol{\theta}} = \left(N^{-1} \sum_{i=1}^{N} \ddot{\mathbf{W}}_i' \ddot{\mathbf{W}}_i\right)^{-1} \left(N^{-1} \sum_{i=1}^{N} \ddot{\mathbf{W}}_i' \mathbf{y}_i\right) = \left(N^{-1} \sum_{i=1}^{N} \ddot{\mathbf{W}}_i' \ddot{\mathbf{W}}_i\right)^{-1} \left[N^{-1} \sum_{i=1}^{N} \ddot{\mathbf{W}}_i' (\mathbf{W}_i \boldsymbol{\theta} + \mathbf{e}_i)\right]$$
$$= \boldsymbol{\theta} + \left(N^{-1} \sum_{i=1}^{N} \ddot{\mathbf{W}}_i' \ddot{\mathbf{W}}_i\right)^{-1} \left(N^{-1} \sum_{i=1}^{N} \ddot{\mathbf{W}}_i' \mathbf{e}_i\right)$$

because $\ddot{\mathbf{W}}_i'\ddot{\mathbf{W}}_i = \ddot{\mathbf{W}}_i'\mathbf{W}_i$.

But

$$N^{-1}\sum_{i=1}^{N}\ddot{\mathbf{W}}_i'\mathbf{e}_i = N^{-1}\sum_{i=1}^{N}\dot{\mathbf{W}}_i'\mathbf{e}_i + N^{-1}\sum_{i=1}^{N}\mathbf{W}_i'(\mathbf{M}_{\hat{\Psi}} - \mathbf{M}_{\Psi})\mathbf{e}_i$$

$$= N^{-1}\sum_{i=1}^{N}\dot{\mathbf{W}}_i'\mathbf{e}_i + \left[N^{-1}\sum_{i=1}^{N}(\mathbf{e}_i' \otimes \mathbf{W}_i')\right]\text{vec}(\mathbf{M}_{\hat{\Psi}} - \mathbf{M}_{\Psi}) \quad (5.21)$$

As we showed earlier, under Assumption 4, $N^{-1}\sum_{i=1}^{N}\dot{\mathbf{W}}_i'\mathbf{e}_i = o_p(1)$. Moreover, we assume $E[(\mathbf{e}_i' \otimes \mathbf{W}_i')]$ exists, and so $N^{-1}\sum_{i=1}^{N}(\mathbf{e}_i' \otimes \mathbf{W}_i') = O_p(1)$. By Assumption 5, $\text{vec}(\mathbf{M}_{\hat{\Psi}} - \mathbf{M}_{\Psi}) = o_p(1)$. We have shown that

$$N^{-1}\sum_{i=1}^{N}\ddot{\mathbf{W}}_i'\mathbf{e}_i = o_p(1) + O_p(1) \cdot o_p(1) = o_p(1) \quad (5.22)$$

and this completes the proof. $\square$

As we saw in Section 3, a leading case of Theorem 2 is

$$\mathbf{M}_{\hat{\Psi}} = \mathbf{M}_{\bar{\mathbf{X}}} = \mathbf{I}_T - \bar{\mathbf{X}}(\bar{\mathbf{X}}'\bar{\mathbf{X}})^{-1}\bar{\mathbf{X}}', \quad (5.23)$$

and necessary for the rank condition is $T > k$. If we add the averages $\bar{\mathbf{y}} = N^{-1}\sum_{i=1}^{N}\mathbf{y}_i$ then necessary for the rank condition is $T > k + 1$.

Theorem 2 applies to CCEP estimators with known deterministic variables, such as

$$\hat{\psi}_t = (1, \bar{\mathbf{x}}_t), \quad (5.24)$$

which seems particularly attractive provided $T > k + 1$. With this choice of $\hat{\psi}_t$ we effectively encompass the usual within estimator and allow much more heterogeneity. Another attractive choice is

$$\hat{\psi}_t = (1, t, \bar{\mathbf{x}}_t), \quad (5.25)$$

which requires $T > k + 2$. In both cases, only some of the elements of $\hat{\Psi}$ are estimated.

In the general case, no restrictions are placed on the covariances between elements of $\mathbf{X}_i$ and $\boldsymbol{\lambda}_i$, which is what gives the analysis a fixed effects flavor. We only require that $\mathbf{X}_i$ is exogenous with respect to the idiosyncratic errors, $\mathbf{e}_i$.

## 6. Consistency with Random Slopes

Without much additional effort, we can extend the consistency result in Section 5 to a model with slope heterogeneity.

**Assumption 1$'$** (**Factor Model**): For a random draw $i$,

$$y_{it} = \mathbf{d}_t \boldsymbol{\alpha}_i + \mathbf{x}_{it} \boldsymbol{\beta}_i + \mathbf{f}_t \boldsymbol{\gamma}_i + e_{it}, \ t = 1, \ldots, T \tag{6.1}$$

where $\mathbf{D}$ ($T \times r$) and $\mathbf{F}$ ($T \times p$) are nonrandom and $\boldsymbol{\alpha}_i$ ($r \times 1$) and $\boldsymbol{\beta}_i$ ($k \times 1$) are random vectors.

☐

In (6.1) both the covariates of interest, $\mathbf{x}_{it}$, and the known deterministic factors $\mathbf{d}_t$ can have heterogeneous coefficients. We can decompose $\boldsymbol{\alpha}_i$ and $\boldsymbol{\beta}_i$ into their means and unit-specific deviations as

$$\boldsymbol{\alpha}_i = \boldsymbol{\alpha} + \mathbf{a}_i, \ \mathrm{E}(\mathbf{a}_i) = \mathbf{0} \tag{6.2}$$
$$\boldsymbol{\beta}_i = \boldsymbol{\beta} + \mathbf{b}_i, \ \mathrm{E}(\mathbf{b}_i) = \mathbf{0} \tag{6.3}$$

and then $\boldsymbol{\alpha}$, and especially $\boldsymbol{\beta}$, become the parameters of interest.

Stacking across $t$ and substituting gives

$$\begin{aligned} \mathbf{y}_i &= \mathbf{D}\boldsymbol{\alpha}_i + \mathbf{X}_i \boldsymbol{\beta}_i + \mathbf{F}\boldsymbol{\gamma}_i + \mathbf{e}_i \\ &= \mathbf{D}\boldsymbol{\alpha} + \mathbf{X}_i \boldsymbol{\beta} + \Psi \boldsymbol{\lambda}_i + \mathbf{e}_i + \mathbf{D}\mathbf{a}_i + \mathbf{X}_i \mathbf{b}_i, \end{aligned} \tag{6.4}$$

where, as before, $\mathbf{F} = \Psi \Lambda$ and $\boldsymbol{\lambda}_i = \Lambda \boldsymbol{\gamma}_i$. Compared with equation (5.5), equation (6.4) has the extra (unobserved) term $\mathbf{D}\mathbf{a}_i + \mathbf{X}_i \mathbf{b}_i$. Because of the small-$T$ framework, we ignore this term in estimation. To ensure we still obtain consistent estimators of the average effects, $\beta$, we add an exogeneity assumption similar to Wooldridge (2005).

**Assumption 6** (**Exogeneity with Respect to Slopes**): With $\mathbf{a}_i$ and $\mathbf{b}_i$ as in (6.2) and (6.3), respectively,

$$\mathrm{E}\left(\dot{\mathbf{X}}_i' \otimes \mathbf{a}_i\right) = \mathrm{E}\left[\left(\sum_{t=1}^{T} \dot{\mathbf{x}}_{it}'\right) \mathbf{a}_i\right] = \mathbf{0} \tag{6.5}$$

$$\mathrm{E}\left(\dot{\mathbf{X}}_i' \otimes \mathbf{b}_i\right) = \mathrm{E}\left[\left(\sum_{t=1}^{T} \dot{\mathbf{x}}_{it}'\right) \mathbf{b}_i\right] = \mathbf{0} \tag{6.6}$$

$$\mathrm{E}\left(\dot{\mathbf{X}}_i' \dot{\mathbf{X}}_i \mathbf{b}_i\right) = \mathrm{E}\left[\left(\sum_{t=1}^{T} \dot{\mathbf{x}}_{it}' \dot{\mathbf{x}}_{it}\right) \mathbf{b}_i\right] = \mathbf{0}. \ \square \tag{6.7}$$

Wooldridge (2005) used a version of Assumption 6 where $\dot{\mathbf{x}}_{it}$ is the unit-specific deviations from the time averages, $\bar{\mathbf{x}}_i = T^{-1} \sum_{t=1}^{T} \mathbf{x}_{it}$, in which case (6.5) and (6.6) require that the unit-specific deviations from means are uncorrelated with $\alpha_i$ and $\beta_i$, respectively. It turns out that (6.6) is not used in the consistency proof, but has a natural interpretation and it implies that a certain composite error term has a mean of zero, which simplifies some calculations in Section 7. Condition (6.7) implies that the unit-specific variances and covariances of $\{\mathbf{x}_{it} : t = 1, \ldots, T\}$ are uncorrelated with $\beta_i$. Wooldridge (2005) also allowed for deviations from unit-specific detrending, where $\mathbf{x}_{it}$ is regressed on, say, $(1, t)$. Here, we can choose $\Psi$ to allow those possibilities while also partialling out of the cross-sectional averages, as is done with CCEP. A related point is that if (6.5) is suspected of being false for some elements of $\mathbf{D}$ then those elements can be moved to $\Psi$ and we net them out of $\mathbf{X}_i$. (As always, this requires $T$ to be sufficiently large.) The idea is that we choose $\Psi$ in a way that includes what we think are the relevant variables to partial out, and then anything omitted from $\Psi$ is in $\mathbf{D}$.

Pesaran (2006) did not study the CCEP estimator with heterogeneous slopes and fixed $T$. Instead, under assumptions much stronger than Assumption 6, he proved that the so-called mean group CCE estimator is consistent with fixed $T$, $N \to \infty$. However, Pesaran assumed, at a minimum, that $\mathbf{x}_{it}$ and $(\boldsymbol{\alpha}_i, \boldsymbol{\beta}_i)$ are independent, so that any systematic heterogeneity in $\mathbf{x}_{it}$ is independent of the heterogeneity on the factors $\mathbf{d}_t$ and the slopes on the covariates of interest. As a simple example of what (6.5) and (6.6) allow when $\boldsymbol{\psi}_t = (1, \boldsymbol{\mu}_t^x)$, suppose $\mathbf{x}_{it} = \mathbf{h}_i + \mathbf{a}_{it}$. Then $\mathbf{h}_i$ can be arbitrarily correlated with $(\boldsymbol{\alpha}_i, \boldsymbol{\beta}_i)$ whereas Pesaran's (2006) analysis of the mean group estimator does not allow that.

Under Assumption 6, the consistency result is a straightforward extension of Theorem 2. As far as we know, this is the first demonstration of consistency of CCEP-type estimators with fixed $T$ and heterogenous slopes – and we need not make strong independence assumptions used to show consistency of other estimators.

**THEOREM 3**: Make the assumptions in Theorem 2 with Assumption $1'$ in place of Assumption 1. In addition, make Assumption 6. Then $\hat{\boldsymbol{\theta}} \overset{p}{\to} \boldsymbol{\theta}$ as $N \to \infty$.

**Proof**: Define a new error term by

$$\mathbf{u}_i \equiv \mathbf{e}_i + \mathbf{D}\mathbf{a}_i + \mathbf{X}_i \mathbf{b}_i \tag{6.8}$$

By Theorem 2, it suffices to show that $E(\dot{\mathbf{u}}_i) = \mathbf{0}$ and $E\left(\dot{\mathbf{X}}_i' \dot{\mathbf{u}}_i\right) = \mathbf{0}$. But

$$E(\dot{\mathbf{u}}_i) = E(\dot{\mathbf{e}}_i) + \dot{\mathbf{D}} E(\mathbf{a}_i) + E(\dot{\mathbf{X}}_i \mathbf{b}_i) = \mathbf{0},$$

and each term is zero by (5.12), the definition of $\mathbf{a}_i$, and (6.6), respectively.

Next,

$$E(\dot{X}'_i\dot{u}_i) = E(\dot{X}'_i\dot{e}_i) + E(\dot{X}'_i\dot{D}a_i) + E(\dot{X}'_i\dot{X}_ib_i) = 0$$

because each term is zero by (5.13), (6.5), and (6.7), respectively. $\square$

# 7. Asymptotic Normality

We now derive the asymptotic normality of the extended CCEP estimator in a setting that allows for constant or random slopes, with and without factors $d_t$. A single theorem applies to all cases. We do not consider the asymptotic distribution of estimators of $\alpha$ but focus on the coefficients $\beta$ of interest. Recall that, in the current setting with $\bar{X}$ included in $\hat{\Psi}$, the estimator $\hat{\beta}$ is the same whether or not we include $D$.

**Assumption 1″ (Linear Model):** For a random draw $i$,

$$y_i = D\alpha + X_i\beta + u_i \tag{7.1}$$

where $y_i$ is $T \times 1$, $X_i$ is $T \times k$, and $u_i$ is $T \times 1$. $\square$

In the model from Section 6,

$$u_i \equiv \Psi\lambda_i + e_i + Da_i + X_ib_i,$$

and in the model of Section 5, $Da_i + X_ib_i$ can be dropped. In this section, we will not assume any special structure for $u_i$.

**Assumption 2′ (Factor Structure):** $\Psi$ is a $T \times m$ matrix with $\mu_X \in \Psi$ and rank$(\Psi) = m$. $\square$

Recall that the matrix $\Psi$ is the population version of the factors we want to remove from $X_i$, at the unit level, in order to consistently estimate $\beta$. Given Assumption 2′, we can define $M_\Psi = I_T - \Psi(\Psi'\Psi)^{-1}\Psi'$, as before. We use the same rank condition as in the consistency result:

**Assumption 3′ (Rank Condition)**: $\mathbf{A} \equiv E(\mathbf{X}_i' \mathbf{M}_\Psi \mathbf{X}_i)$ is nonsingular. □

We state the exogeneity assumption in a way that covers the previous examples while allowing a simple, general derivation of asymptotic normality:

**Assumption 4′ (Exogeneity)**: With $\dot{\mathbf{u}}_i = \mathbf{M}_\Psi \mathbf{u}_i$ and $\dot{\mathbf{X}}_i = \mathbf{M}_\Psi \mathbf{X}_i$,

$$E(\dot{\mathbf{u}}_i) = \mathbf{0} \tag{7.2}$$

and

$$E(\dot{\mathbf{X}}_i' \mathbf{u}_i) = E(\dot{\mathbf{X}}_i' \dot{\mathbf{u}}_i) = \mathbf{0}. \ \square \tag{7.3}$$

Assumption 4′ holds for the models in Section 5 (under Assumption 4) and Section 6 (under Assumptions 4 and 6). While $E(\dot{\mathbf{u}}_i) = \mathbf{0}$ is not necessary for consistency or asymptotic normality, it does simplify the first-order representation $\sqrt{N}(\hat{\boldsymbol{\beta}} - \boldsymbol{\beta})$ in a useful way. [By contrast, $E(\mathbf{u}_i) = \mathbf{0}$ is too strong and does not hold in the settings of Sections 5 and 6.]

**Assumption 5′ (Asymptotic Normality of $\hat{\Psi}$)**: For an estimator $\hat{\Psi}$ that includes $\bar{\mathbf{X}}$ in its columns,

$$\text{vec}\left[\sqrt{N}(\hat{\Psi} - \Psi)\right] = N^{-1/2} \sum_{i=1}^{N} \mathbf{q}_i(\Psi) + o_p(1) \tag{7.4}$$

where

$$E[\mathbf{q}_i(\Psi)] = \mathbf{0}. \ \square \tag{7.5}$$

Naturally, Assumption 5′ implies Assumption 5. Define the estimator of $\boldsymbol{\beta}$ as

$$\hat{\boldsymbol{\beta}} = \left(\sum_{i=1}^{N} \ddot{\mathbf{X}}_i' \ddot{\mathbf{X}}_i\right)^{-1} \left(\sum_{i=1}^{N} \ddot{\mathbf{X}}_i' \mathbf{y}_i\right),$$

where $\ddot{\mathbf{X}}_i$ is the $T \times k$ matrix of residuals

$$\ddot{\mathbf{X}}_i = \mathbf{M}_{\hat{\Psi}}\mathbf{X}_i.$$

To derive asymptotic normality, note that $\sum_{i=1}^{N} \ddot{\mathbf{X}}_i'\mathbf{D} = \mathbf{0}$ and so

$$\sqrt{N}\left(\hat{\boldsymbol{\beta}} - \boldsymbol{\beta}\right) = \left(N^{-1}\sum_{i=1}^{N} \ddot{\mathbf{X}}_i'\ddot{\mathbf{X}}_i\right)^{-1}\left(N^{-1/2}\sum_{i=1}^{N} \ddot{\mathbf{X}}_i'\mathbf{u}_i\right). \qquad (7.6)$$

From the consistency proof, we already know that

$$N^{-1}\sum_{i=1}^{N} \ddot{\mathbf{X}}_i'\ddot{\mathbf{X}}_i = \mathbf{A} + o_p(1)$$

$$\mathbf{A} \equiv E(\mathbf{X}_i'\mathbf{M}_{\Psi}\mathbf{X}_i) \qquad (7.7)$$

Along the way, we will show that

$$N^{-1/2}\sum_{i=1}^{N} \ddot{\mathbf{X}}_i'\mathbf{u}_i = O_p(1)$$

and so, by the asymptotic equivalence lemma,

$$\sqrt{N}\left(\hat{\boldsymbol{\beta}} - \boldsymbol{\beta}\right) = \mathbf{A}^{-1}\left(N^{-1/2}\sum_{i=1}^{N} \ddot{\mathbf{X}}_i'\mathbf{u}_i\right) + o_p(1) \qquad (7.8)$$

Therefore, as usual in these contexts, we can replace the average $N^{-1}\sum_{i=1}^{N} \ddot{\mathbf{X}}_i'\ddot{\mathbf{X}}_i$ with its nonrandom probability limit without affecting the limiting distribution.

Unless we make additional assumptions, the asymptotic distribution of $N^{-1/2}\sum_{i=1}^{N} \ddot{\mathbf{X}}_i'\mathbf{u}_i$ is not the same as that of $N^{-1/2}\sum_{i=1}^{N} \dot{\mathbf{X}}_i'\mathbf{u}_i$. In other words, estimation of $\Psi$ generally affects the asymptotic distribution of $\sqrt{N}\left(\hat{\boldsymbol{\beta}} - \boldsymbol{\beta}\right)$. To see this, write

$$N^{-1/2}\sum_{i=1}^{N} \ddot{\mathbf{X}}_i'\mathbf{u}_i = N^{-1/2}\sum_{i=1}^{N} (\mathbf{X}_i - \bar{\mathbf{X}})'\mathbf{M}_{\hat{\Psi}}\mathbf{u}_i,$$

which follows from $\mathbf{M}'_{\hat{\Psi}}\bar{\mathbf{X}} = \mathbf{0}$. Next, write

$$N^{-1/2}\sum_{i=1}^{N}(\mathbf{X}_i - \bar{\mathbf{X}})'\mathbf{M}_{\hat{\Psi}}\mathbf{u}_i = N^{-1/2}\sum_{i=1}^{N}(\mathbf{X}_i - \bar{\mathbf{X}})'\mathbf{M}_{\Psi}\mathbf{u}_i + N^{-1/2}\sum_{i=1}^{N}(\mathbf{X}_i - \bar{\mathbf{X}})'(\mathbf{M}_{\hat{\Psi}} - \mathbf{M}_{\Psi})\mathbf{u}_i$$

$$= N^{-1/2}\sum_{i=1}^{N}\dot{\mathbf{X}}'_i\mathbf{u}_i - N^{-1/2}\sum_{i=1}^{N}\bar{\mathbf{X}}'\mathbf{M}_{\Psi}\mathbf{u}_i \qquad (7.9)$$

$$+ N^{-1/2}\sum_{i=1}^{N}(\mathbf{X}_i - \bar{\mathbf{X}})'(\mathbf{M}_{\hat{\Psi}} - \mathbf{M}_{\Psi})\mathbf{u}_i$$

The middle term can be written as

$$N^{-1/2}\sum_{i=1}^{N}\bar{\mathbf{X}}'\mathbf{M}_{\Psi}\mathbf{u}_i = \bar{\mathbf{X}}'\mathbf{M}_{\Psi}\left(N^{-1/2}\sum_{i=1}^{N}\mathbf{M}_{\Psi}\mathbf{u}_i\right) = \bar{\mathbf{X}}'\mathbf{M}_{\Psi}\left(N^{-1/2}\sum_{i=1}^{N}\dot{\mathbf{u}}_i\right) \qquad (7.10)$$

By (7.2) and the CLT, $N^{-1/2}\sum_{i=1}^{N}\dot{\mathbf{u}}_i = O_p(1)$. By the LLN and Slutsky's Theorem,

$\bar{\mathbf{X}}'\mathbf{M}_{\Psi} \xrightarrow{p} \boldsymbol{\mu}'_{\mathbf{X}}\mathbf{M}_{\Psi} = \mathbf{0}$. Therefore, the middle term is $o_p(1)$. For the final term, write it as

$$N^{-1/2}\sum_{i=1}^{N}(\mathbf{X}_i - \bar{\mathbf{X}})'(\mathbf{M}_{\hat{\Psi}} - \mathbf{M}_{\Psi})\mathbf{u}_i = \left\{N^{-1}\sum_{i=1}^{N}[\mathbf{u}_i \otimes (\mathbf{X}_i - \bar{\mathbf{X}})]\right\}' \text{vec}\left[\sqrt{N}(\mathbf{M}_{\hat{\Psi}} - \mathbf{M}_{\Psi})\right] \qquad (7.11)$$

Now

$$\text{vec}\left[\sqrt{N}(\mathbf{M}_{\hat{\Psi}} - \mathbf{M}_{\Psi})\right] = O_p(1)$$

and

$$N^{-1}\sum_{i=1}^{N}[\mathbf{u}_i \otimes (\mathbf{X}_i - \bar{\mathbf{X}})] = N^{-1}\sum_{i=1}^{N}[\mathbf{u}_i \otimes (\mathbf{X}_i - \boldsymbol{\mu}_{\mathbf{X}})] + \left(N^{-1}\sum_{i=1}^{N}\mathbf{u}_i\right) \otimes (\boldsymbol{\mu}_{\mathbf{X}} - \bar{\mathbf{X}})$$

$$= N^{-1}\sum_{i=1}^{N}[\mathbf{u}_i \otimes (\mathbf{X}_i - \boldsymbol{\mu}_{\mathbf{X}})] + O_p(1) \cdot o_p(1)$$

So we have shown

$$N^{-1/2} \sum_{i=1}^{N} \ddot{\mathbf{X}}_i' \mathbf{u}_i = N^{-1/2} \sum_{i=1}^{N} \dot{\mathbf{X}}_i' \mathbf{u}_i + \left\{ N^{-1} \sum_{i=1}^{N} [\mathbf{u}_i \otimes (\mathbf{X}_i - \boldsymbol{\mu}_\mathbf{X})] \right\}' \text{vec}\left[ \sqrt{N} (\mathbf{M}_{\hat{\boldsymbol{\Psi}}} - \mathbf{M}_{\boldsymbol{\Psi}}) \right] + o_p(1) \quad (7.12)$$

and by the LLN,

$$N^{-1/2} \sum_{i=1}^{N} \ddot{\mathbf{X}}_i' \mathbf{u}_i = N^{-1/2} \sum_{i=1}^{N} \dot{\mathbf{X}}_i' \mathbf{u}_i + \mathbf{G}' \text{vec}\left[ \sqrt{N} (\mathbf{M}_{\hat{\boldsymbol{\Psi}}} - \mathbf{M}_{\boldsymbol{\Psi}}) \right] + o_p(1) \quad (7.13)$$

where

$$\mathbf{G} \equiv E[\mathbf{u}_i \otimes (\mathbf{X}_i - \boldsymbol{\mu}_\mathbf{X})] \quad (7.14)$$

is $\mathbf{G}$ is $T^2 \times k$. From (7.13) and (7.14), the asymptotic distribution of $\text{vec}\left[ \sqrt{N} (\mathbf{M}_{\hat{\boldsymbol{\Psi}}} - \mathbf{M}_{\boldsymbol{\Psi}}) \right]$ does not affect the asymptotic distribution of $\sqrt{N} (\hat{\boldsymbol{\beta}} - \boldsymbol{\beta})$ if each element of $\mathbf{X}_i$ is uncorrelated with each element of $\mathbf{u}_i$. This is a strong requirement. For example, in the basic constant coefficient model in equation (7.2), we would require $E[(\mathbf{X}_i - \boldsymbol{\mu}_\mathbf{X}) \otimes \boldsymbol{\gamma}_i] = \mathbf{0}$, which means that each element of $\mathbf{X}_i$ is uncorrelated with the heterogeneity. This is tantamount to a standard "random effects" assumption and is needed neither for consistency nor asymptotic normality.

The next step is to apply the delta method to find the asymptotic variance of $\text{vec}\left[ \sqrt{N} (\mathbf{M}_{\hat{\boldsymbol{\Psi}}} - \mathbf{M}_{\boldsymbol{\Psi}}) \right]$. According to Abadir and Magnus (2005, Exercise 13.24), the Jacobian of $\text{vec}(\mathbf{M}_{\boldsymbol{\Psi}})$ with respect to $\text{vec}(\boldsymbol{\Psi})$ is

$$(\mathbf{I}_{T^2} + \mathbf{K}_{T^2})\left[ \boldsymbol{\Psi}(\boldsymbol{\Psi}'\boldsymbol{\Psi})^{-1} \otimes \mathbf{M}_{\boldsymbol{\Psi}} \right], \quad (7.15)$$

where $\mathbf{K}_{T^2}$ is the $T^2 \times T^2$ commutation matrix [see Abadir and Magnus (2005, page 299)]. It follows from (7.13), (7.15), and the delta method that

$$N^{-1/2} \sum_{i=1}^{N} \mathbf{X}_i'(\mathbf{M}_{\hat{\boldsymbol{\Psi}}} - \mathbf{M}_{\boldsymbol{\Psi}})\mathbf{u}_i = \mathbf{G}'(\mathbf{I}_{T^2} + \mathbf{K}_{T^2})\left[ \boldsymbol{\Psi}(\boldsymbol{\Psi}'\boldsymbol{\Psi})^{-1} \otimes \mathbf{M}_{\boldsymbol{\Psi}} \right]\text{vec}\left[ \sqrt{N} (\hat{\boldsymbol{\Psi}} - \boldsymbol{\Psi}) \right] + o_p(1) \quad (7.16)$$

For the basic CCEP estimator without deterministic factors in $\mathbf{x}_{it}$,

$$\hat{\boldsymbol{\Psi}} = N^{-1} \sum_{i=1}^{N} \mathbf{X}_i = \bar{\mathbf{X}},$$

and then the CLT applies directly. If we want to remove unit-specific averages or time trends, some elements of $\hat{\boldsymbol{\Psi}}$ are not random. Generally, assume the representation (7.4). The variance-covariance matrix of $\mathbf{q}_i(\boldsymbol{\Psi})$ may be singular, as occurs when some elements of $\boldsymbol{\Psi}$ are not estimated, but this causes no difficulties. Dropping the dependence of $\mathbf{q}_i(\boldsymbol{\Psi})$ on $\boldsymbol{\Psi}$, define

$$\mathbf{s}_i \equiv \dot{\mathbf{X}}_i' \mathbf{u}_i + \mathbf{G}'(\mathbf{I}_{T^2} + \mathbf{K}_{T^2})\left[\boldsymbol{\Psi}(\boldsymbol{\Psi}'\boldsymbol{\Psi})^{-1} \otimes \mathbf{M}_{\boldsymbol{\Psi}}\right]\mathbf{q}_i \qquad (7.17)$$

Under Assumptions 1′ through 5′ – whether or not we have random slopes – $\mathrm{E}(\mathbf{s}_i) = \mathbf{0}$, and so we can apply the CLT:

$$N^{-1/2} \sum_{i=1}^{N} \mathbf{s}_i \xrightarrow{d} \mathrm{Normal}(\mathbf{0}, \mathbf{B}) \qquad (7.18)$$

$$\mathbf{B} \equiv \mathrm{Var}(\mathbf{s}_i) = \mathrm{E}(\mathbf{s}_i \mathbf{s}_i') \qquad (7.19)$$

We now have the representation

$$\sqrt{N}\left(\hat{\boldsymbol{\beta}} - \boldsymbol{\beta}\right) = \mathbf{A}^{-1}\left(N^{-1/2} \sum_{i=1}^{N} \mathbf{s}_i\right) + o_p(1) \qquad (7.20)$$

and so, by the asymptotic equivalence lemma,

$$\sqrt{N}\left(\hat{\boldsymbol{\beta}} - \boldsymbol{\beta}\right) \xrightarrow{d} \mathrm{Normal}(\mathbf{0}, \mathbf{A}^{-1}\mathbf{B}\mathbf{A}^{-1}) \qquad (7.21)$$

We have proven the following result.

**THEOREM 4**: (i) Under Assumption 1″ and Assumptions 2′ through 5′ (and standard regularity conditions), (7.21) holds with $\mathbf{B}$ given by (7.19) and $\mathbf{A}$ given in (7.7). (ii) If, in addition,

$$E[\mathbf{u}_i \otimes (\mathbf{X}_i - \boldsymbol{\mu}_\mathbf{X})] = \mathbf{0} \tag{7.22}$$

then $\mathrm{AVar}\sqrt{N}\left(\hat{\boldsymbol{\beta}} - \boldsymbol{\beta}\right)$ does not depend on the asymptotic variance of $\mathrm{vec}\left[\sqrt{N}\left(\hat{\boldsymbol{\Psi}} - \boldsymbol{\Psi}\right)\right]$. □

Theorem 4 considerably expands the scope of CCEP-type estimators to include the two commonly used versions of CCEP as well as hybrids of standard fixed effects estimators and CCEP. The result allows for substantial heterogeneity correlated with $\mathbf{x}_{it}$ provided the choice of $\boldsymbol{\Psi}$ eliminates the heterogeneity. Arbitary serial correlation and heteroskedasticity is allowed in $\{u_{it} : t = 1, \ldots, T\}$, allowing for the standard model with fixed coefficients and serially correlated $\{e_{it} : t = 1, \ldots, T\}$ as well as the random slopes models discussed in Section 6.

When applied to the special case of the CCEP($\bar{\mathbf{Z}}$) estimator, (7.21) differs from that in WPN (2019) because the latter does not account for sampling variation in $\bar{\mathbf{Z}}$. Here, we explicitly account for the first-stage estimation error. Assumption (7.22) then provides the strong condition under which that sampling variation can be ignored.

## 8. Estimating the Asymptotic Variance

A consistent estimator of

$$\mathrm{AVar}\left[\sqrt{N}\left(\hat{\boldsymbol{\beta}} - \boldsymbol{\beta}\right)\right] = \mathbf{A}^{-1}\mathbf{B}\mathbf{A}^{-1} \tag{8.1}$$

is obtained by consistently estimating $\mathbf{A}$ and $\mathbf{B}$. We have a consistent estimator of $\mathbf{A}$ under the assumptions of Theorem 4:

$$\hat{\mathbf{A}} = N^{-1} \sum_{i=1}^{N} \ddot{\mathbf{X}}_i' \ddot{\mathbf{X}}_i = N^{-1} \sum_{i=1}^{N} \mathbf{X}_i' \mathbf{M}_{\hat{\boldsymbol{\Psi}}} \mathbf{X}_i. \tag{8.2}$$

For $\mathbf{B}$, we need to estimate the variance of $\mathbf{s}_i$ in (7.17). Under weak regularity conditions – we will not state a formal result because the conditions are standard extensions of the assumptions in Theorem 4 – we can replace all unknown parameters with consistent estimators and replace population means with sample averages. In obtaining the residuals, there is a somewhat subtle issue. Namely, the expression for $\mathbf{B}$ in (7.19) we should obtain residuals that correspond to the model, (7.1). To this end, generally define

$$\hat{\mathbf{u}}_i = \mathbf{y}_i - \mathbf{D}\hat{\boldsymbol{\alpha}} - \mathbf{X}_i\hat{\boldsymbol{\beta}}, \tag{8.3}$$

so that the residuals net out $\mathbf{D}$ as well as $\mathbf{X}_i$. From Theorem 1, if $\bar{\mathbf{y}} \in \hat{\boldsymbol{\Psi}}$ along with $\bar{\mathbf{X}}$, $\hat{\boldsymbol{\alpha}} = \mathbf{0}$. In most cases it will be true that the sample analog of (7.2) holds: $N^{-1}\sum_{i=1}^{N}\mathbf{M}_{\hat{\boldsymbol{\Psi}}}\hat{\mathbf{u}}_i = \mathbf{0}$.

With these residuals, define

$$\hat{\mathbf{G}} = N^{-1}\sum_{i=1}^{N}[\hat{\mathbf{u}}_i \otimes (\mathbf{X}_i - \bar{\mathbf{X}})] \tag{8.4}$$

$$\hat{\mathbf{q}}_i = \mathbf{q}_i(\hat{\boldsymbol{\Psi}}) \tag{8.5}$$

Next, define the $k \times 1$ vector

$$\hat{\mathbf{s}}_i \equiv \ddot{\mathbf{X}}_i'\hat{\mathbf{u}}_i + \hat{\mathbf{G}}'(\mathbf{I}_{T^2} + \mathbf{K}_{T^2})\left[\hat{\boldsymbol{\Psi}}\left(\hat{\boldsymbol{\Psi}}'\hat{\boldsymbol{\Psi}}\right)^{-1} \otimes \mathbf{M}_{\hat{\boldsymbol{\Psi}}}\right]\hat{\mathbf{q}}_i \tag{8.6}$$

and the $k \times k$ matrix

$$\hat{\mathbf{B}} = N^{-1}\sum_{i=1}^{N}\hat{\mathbf{s}}_i\hat{\mathbf{s}}_i' \tag{8.7}$$

From (8.1) a consistent estimator of $\mathrm{AVar}\left[\sqrt{N}\left(\hat{\boldsymbol{\beta}} - \boldsymbol{\beta}\right)\right]$ is $\hat{\mathbf{A}}^{-1}\hat{\mathbf{B}}\hat{\mathbf{A}}^{-1}$.

For the CCEP estimators we have

$$\hat{\mathbf{q}}_i = \mathrm{vec}(\mathbf{X}_i - \bar{\mathbf{X}})$$

or

$$\hat{\mathbf{q}}_i = \mathrm{vec}(\mathbf{X}_i - \bar{\mathbf{X}}, \mathbf{y}_i - \bar{\mathbf{y}})$$

If $\psi_t$ contains known elements, such as unity or $(1,t)$, then some of the elements of $\hat{\mathbf{q}}_i$ are identically zero.

## 9. Practical Considerations

Given the widespread popularity of the TWFE estimator in the microeconometrics literature and the popularity of CCEP in the large-$T$ literature, it seems natural, with enough time periods, to have the best of both worlds and use the estimator that nets out

$$\hat{\psi}_t \equiv (1, \bar{\mathbf{x}}_t)$$

from $\{\mathbf{x}_{it} : t = 1, \ldots, T\}$, unit by unit. Recall from Section 3 that including $\bar{\mathbf{x}}_t$ makes the inclusion of any aggregate variables $\mathbf{d}_t$ – specifically, time dummies or observed macro variables with constant coefficients – irrelevant for obtaining $\hat{\boldsymbol{\beta}}$. Given that $\hat{\psi}_t$ has $k + 1$ elements a requirement on the number of time periods is $T > k + 1$. (The usual TWFE estimator requires only $T \geq 2$.) There is no basis for thinking that the matrix $(\mathbf{j}_T, \boldsymbol{\mu}_X)$, where $\mathbf{j}_T$ is the $T \times 1$ vector of ones, has rank less than $k + 1$, and so we are comfortable imposing Assumption 2'. Also, under the assumptions in Sections 6 and 7, the underlying model can have heterogeneity in the coefficients on $\mathbf{x}_{it}$ and on deterministic variables, $\mathbf{d}_t$, and we still consistently estimate the average partial effects, $\boldsymbol{\beta} = \mathrm{E}(\boldsymbol{\beta}_i)$. For obtaining the asymptotic variance estimator of $\hat{\boldsymbol{\beta}}$, as many aggregate time dummies as allowable should be included to ensure the residuals have, in sample, the same property as (7.2) in the population.

A popular "robustness check" in many panel data applications is to allow unit-specific linear trends, in which case $t$ gets added to $\hat{\psi}_t$. Now we require $T > k + 2$ but, if the time periods are available, partialling out $(1, t, \bar{\mathbf{x}}_t)$ sets a pretty high bar for estimation of $\boldsymbol{\beta}$.

With more time periods one can partial out $(1, \bar{\mathbf{x}}_t, \bar{y}_t)$ or even $(1, t, \bar{\mathbf{x}}_t, \bar{y}_t)$ in the extended

CCEP estimator. One should expect the precision of $\hat{\boldsymbol{\beta}}$ to suffer because even more variation is removed from $\{\mathbf{x}_{it} : t = 1,\ldots,T\}$. Adding $\bar{\mathbf{y}}$ along with $\bar{\mathbf{X}}$ is allowed provided $\boldsymbol{\mu}_y$ is not perfectly collinear with $\boldsymbol{\mu}_X$. We showed in Section 4 that in the traditional CCE framework, when $p = k + 1$ and we use $(\bar{\mathbf{X}}, \bar{\mathbf{y}})$ in the CCEP estimation, the estimator is consistent. [This is also a special case of Theorem 1, which is derived under much weaker assumptions than WPN (2019).]

There is one case that falls through the cracks of our framework. Namely, when we use CCEP$(\bar{\mathbf{X}}, \bar{\mathbf{y}})$ (or its extensions) and the rank condition fails. In the WPN (2019) setting, this can only happen when $p < k + 1$ and there are no deterministic variables in the main equation. Then

$$\boldsymbol{\mu}_y = \boldsymbol{\mu}_X \boldsymbol{\beta} + \mathbf{F}\boldsymbol{\gamma}$$

and, along with (2.7),

$$\boldsymbol{\mu}_y = \boldsymbol{\mu}_X \boldsymbol{\beta} + \boldsymbol{\mu}_X \boldsymbol{\Lambda}_X \boldsymbol{\gamma} = \boldsymbol{\mu}_X (\boldsymbol{\beta} + \boldsymbol{\Lambda}_X \boldsymbol{\gamma}),$$

which shows that $\boldsymbol{\mu}_y$ is a linear function of $\boldsymbol{\mu}_X$. The link between $\boldsymbol{\mu}_y$ and $\boldsymbol{\mu}_X$ is broken if deterministic factors appear in the main equation or if there is heterogeneity in the slopes, $\boldsymbol{\beta}_i$. Our view is that situations where one chooses to include $\bar{\mathbf{y}}$ while wanting to allow rank deficiency are relatively unimportant.

If we take the empirically relevant case where $(\boldsymbol{\mu}_X, \boldsymbol{\mu}_y)$ has rank $k + 1$ (full rank), then we can show, under some strong assumptions, that including $\bar{\mathbf{y}}$ generally increases the asymptotic variance. Suppose we start with the equation

$$\mathbf{y}_i = \mathbf{D}\boldsymbol{\alpha} + \mathbf{X}_i \boldsymbol{\beta} + \mathbf{u}_i$$

Assume $E(\mathbf{M}_{\boldsymbol{\mu}_Z}\mathbf{u}_i) = \mathbf{0}$ and $E[\mathbf{u}_i \otimes (\mathbf{X}_i - \boldsymbol{\mu}_X)] = \mathbf{0}$, so that estimation of $\boldsymbol{\mu}_X$ and $\boldsymbol{\mu}_y$ does not affect the asymptotic distribution of $CCEP(\bar{\mathbf{X}})$ or $CCEP(\bar{\mathbf{X}}, \bar{\mathbf{y}})$ (Theorem 4). Now add an "ideal" set of assumptions that rules out general heteroskedasticity and serial correlation in $\{u_{it} : t = 1, \ldots, T\}$:

$$E(\dot{\mathbf{X}}_i' \mathbf{u}_i \mathbf{u}_i' \dot{\mathbf{X}}_i) = \sigma_u^2 E(\dot{\mathbf{X}}_i' \dot{\mathbf{X}}_i)$$
$$E(\check{\mathbf{X}}_i' \mathbf{u}_i \mathbf{u}_i' \check{\mathbf{X}}_i) = \sigma_u^2 E(\check{\mathbf{X}}_i' \check{\mathbf{X}}_i)$$

where $\dot{\mathbf{X}}_i = \mathbf{M}_{\boldsymbol{\mu}_X}\mathbf{X}_i$ and $\check{\mathbf{X}}_i = \mathbf{M}_{\boldsymbol{\mu}_Z}\mathbf{X}_i$. Applying Theorem 3, it is straightforward to show that

$$\mathrm{AVar}\left[\sqrt{N}\left(\hat{\boldsymbol{\beta}}_{CCEP(\bar{\mathbf{X}})} - \boldsymbol{\beta}\right)\right] = \sigma_u^2 \left[E(\dot{\mathbf{X}}_i'\dot{\mathbf{X}}_i)\right]^{-1} \tag{9.1}$$

and

$$\mathrm{AVar}\left[\sqrt{N}\left(\hat{\boldsymbol{\beta}}_{CCEP(\bar{\mathbf{X}},\bar{\mathbf{y}})} - \boldsymbol{\beta}\right)\right] = \sigma_u^2 \left[E(\check{\mathbf{X}}_i'\check{\mathbf{X}}_i)\right]^{-1} \tag{9.2}$$

Furthermore, because $\check{\mathbf{X}}_i$ nets out $\boldsymbol{\mu}_Z$ from $\mathbf{X}_i$ whereas $\dot{\mathbf{X}}_i$ nets out only $\boldsymbol{\mu}_X$, $E(\dot{\mathbf{X}}_i'\dot{\mathbf{X}}_i) - E(\check{\mathbf{X}}_i'\check{\mathbf{X}}_i)$ is positive semi-definite, which in turn implies

$$\left[E(\check{\mathbf{X}}_i'\check{\mathbf{X}}_i)\right]^{-1} - \left[E(\dot{\mathbf{X}}_i'\dot{\mathbf{X}}_i)\right]^{-1}$$

is PSD. In other words, under an ideal set of assumptions $CCEP(\bar{\mathbf{X}})$ is asymptotically more efficient than $CCEP(\bar{\mathbf{X}}, \bar{\mathbf{y}})$. Because this efficiency claim holds only under strong assumptions, we prefer to argue for $\hat{\boldsymbol{\beta}}_{CCEP(\bar{\mathbf{X}})}$, and extended versions, based mainly on its being more natural and already allowing much more heterogeneity than usual.

## 10. Concluding Remarks

We have proposed an alternative formulation of factor models in the case of small-$T$ panel data settings. In the simplest case, we assume that the unobserved factors, $\mathbf{F}$, can be expressed as linear functions of $\mu_X$, the means of the explanatory variables. Our framework is more natural and, in relevant cases, the assumptions are more general than currently available frameworks. Plus, we show how to extend the basic CCEP estimator to account for staples of applied microeconomics such as heterogenous intercepts and trends. Estimation is straightforward, and we prove consistency under weak assumptions. Idiosyncratic errors are allowed to be arbitrarily serially correlated and both idiosyncratic errors and unobserved heterogeneity can be arbitrarily heteroskedasticity. We also provide sufficient conditions under which the presence of heterogeneous slopes does not affect consistency of the (extended) CCEP estimator of the average partial effects. Our asymptotic normality result applies to a broad class of extended CCEP estimators, and our proposed estimator of the asymptotic variance is straightforward.

Further work could relax the strict exogeneity assumption on the explanatory variables, although allowing for a lot of unobserved heterogeneity makes that assumption more realistic. Nevertheless, the methods in this paper do not apply to models with lagged dependent variables or other situations where shocks today affect explanatory variables either contemporaneously or with a lag.